\documentclass[aps,prb,twocolumn,showpacs,preprintnumbers,amsmath,amssymb,superscriptaddress,longbibliography]{revtex4-2}

\usepackage [latin1]{inputenc}
\usepackage{graphicx}
\usepackage{dcolumn}
\usepackage{bm}
\usepackage{color}
\usepackage[bookmarks=false,linkcolor=blue,urlcolor=blue,colorlinks,citecolor=blue]{hyperref}
\newcommand{\nc}{\newcommand}
\nc{\be}{\begin{equation}}
\nc{\ee}{\end{equation}}
\nc{\bea}{\begin{eqnarray}}
\nc{\eea}{\end{eqnarray}}
\nc{\bean}{\begin{eqnarray*}}
\nc{\eean}{\end{eqnarray*}}
\nc{\mb}{\mbox}
\nc{\rnc}{\renewcommand}
\nc{\vk}{\mb{\bf k}}
\nc{\vp}{\mb{\bf p}}
\nc{\vn}{\mb{\bf n}}
\nc{\vq}{\mb{\bf q}}
\nc{\rr}{\mb{\bf r}}
\nc{\vz}{\hat {\mb{\bf z}}}
\nc{\vj}{\mb{\boldmath$j$}}
\nc{\vg}{\mb{\boldmath$g$}}
\nc{\x}{\mb{\boldmath$x$}}
\nc{\A}{\mb{\boldmath$A$}}
\nc{\va}{\mb{\boldmath$a$}}
\nc{\vs}{\mb{\boldmath$\sigma$}}
\nc{\vpi}{\mb{\boldmath$\pi$}}
\nc{\nab}{\nabla}
\nc{\X}{\sf x}
\nc{\kvec}{\mathbf{k}}
\nc{\pvec}{\mathbf{p}}
\nc{\qvec}{\mathbf{q}}
\nc{\kk}{{\bf k}}
\nc{\pp}{{\bf p}}
\nc{\qq}{{\bf q}}
\nc{\upspin}{{\uparrow}}
\nc{\dspin}{{\downarrow}}

\begin{document}

\title{Effect of Inversion Asymmetry on Bilayer Graphene's Superconducting and Exciton Condensates}

\author{Xiang Hu}
\affiliation{School of Physical Science and Technology, Guangxi Normal University, Guilin 541004, China}
\author{Enrico Rossi}
\affiliation{Department of Physics, William \& Mary, Williamsburg, VA, 23187.}
\author{Yafis Barlas}
\affiliation{Department of Physics, University of Nevada, Reno, NV, 89557.}

\date{\today}
\begin{abstract}
Inversion asymmetry in bilayer graphene can be tuned by the displacement field. As a result, the band dispersion in biased bilayer graphene acquires flat band regions near the Dirac points along with a non-trivial band geometry. We analyze the effect of inversion asymmetry on the critical temperature and superfluid stiffness of the superconducting state of AB-stacked graphene bilayer and the exciton condensate in double layers formed by two AB-stacked graphene bilayers.
We find that the geometric superfluid stiffness in bilayer graphene superconductors is negligible due to the small superconducting gap. Furthermore, since the geometric superfluid stiffness is maximized for a constant order parameter, it can be neglected in biased bilayer graphene superconductors with any pairing symmetry. In contrast, the displacement field enhances the geometric superfluid stiffness in exciton condensates. It is most prominent at low densities and high displacement fields. A consequence of the geometric superfluid stiffness is a modest enhancement of the Berezinskii-Kosterlitz-Thouless transition temperature in bilayer graphene's exciton condensate.
\end{abstract}

\maketitle

\section{Introduction}
The recent discovery of superconductivity~\cite{Cao2018,Yankowitz2019,Cao2018second,Lu2019,Chen2019,Park2021,Cao2021} and correlated phases~\cite{Sharpe605,PhysRevLett.123.197702,Cao2020,Polshyn_2020,Chen2021,Xie2021} in twisted two-dimensional crystals~\cite{bistritzer2011moire} has brought attention to the role of non-trivial band geometry in multi-orbital superconductors~\cite{Peotta2015,PhysRevB.95.024515,PhysRevLett.123.237002,PhysRevLett.124.167002,PhysRevB.101.060505,PhysRevLett.127.170404,PhysRevB.98.220511,PhysRevB.102.184504,PhysRevLett.117.045303,Rossi2021quantum} and other strongly correlated states.~\cite{Roysondhi2013816,PhysRevB.90.165139,PhysRevLett.107.116801,PhysRevB.85.241308,PhysRevX.1.021014,Jackson2015,Xie2021,PhysRevB.101.235312,PhysRevResearch.2.023238,PhysRevResearch.2.023237,PhysRevB.103.125406,PhysRevLett.126.137601,PhysRevB.105.L140506} Non-trivial band geometry in multi-orbital superconductors~\cite{Cao2018,Yankowitz2019,Cao2018second,Lu2019,Chen2019,Park2021,Cao2021} and exciton condensates results in a geometric superfluid stiffness associated with inter-band excitations of the condensate. For isolated bands, the interband contribution to the superfluid stiffness can be projected onto the lowest energy band and is proportional to the quantum metric of that band~\cite{PhysRevB.95.024515}. In contrast, the conventional superfluid stiffness is proportional to the electron density and inversely proportional to the band's effective mass.~\cite{schrieffer1999theory} Consequently, the geometric superfluid weight dominates in flat band superconductors~\cite{PhysRevLett.123.237002,PhysRevLett.124.167002} and exciton condensates~\cite{PhysRevLett.126.137601,Wang2019,PhysRevB.105.L140506}, such as superconducting twisted bilayer graphene~\cite{Cao2018,Yankowitz2019,Cao2018second,Lu2019} and twisted multi-layer graphene~\cite{Chen2019,Park2021,Cao2021} at magic angles. Additionally, the quantum metric is lower-bounded by the absolute value of the Berry curvature. Therefore 
the superconducting states of systems
with topological and Wannier obstructed bands~\cite{Peotta2015,PhysRevLett.128.087002,PhysRevLett.124.167002,PhysRevB.94.254149,PhysRevB.102.201112} 
are guaranteed to have a nonzero geometric contribution to the superfluid stiffness~\cite{Tian2023}.

Until recently, the geometric superfluid weight has only been studied for flat or weakly dispersive isolated bands. However, situations arise where an otherwise dispersive band contains large flat regions in momentum space, as in multi-layer graphene systems~\cite{PhysRevLett.96.086805,PhysRevB.77.155416,PhysRevB.73.245426,PhysRevB.81.115315,PhysRevB.87.115422,PhysRevB.83.165443}. Since pairing interactions are typically projected close to the Fermi energy if the Fermi energy is within these flat regions, the geometric superfluid density can be comparable to its conventional counterpart. Additionally, to maximize the flat band regions and geometric superfluid stiffness, it would be ideal to engineer situations where the local extrema of the Berry curvature also coincides with such flat band regions. In bilayer graphene, the flat band regions appear at the Dirac points, in the vicinity of which the Berry curvature and the quantum metric exhibit maximum values. The tunability of the band structure and Berry curvature of biased bilayer graphene (BLG) by displacement fields, and the control of the total density, in dual-gated samples make it possible to satisfy these stringent constraints.  

In this work, we analyze the geometric and conventional superfluid stiffness ratio, which can be tuned by displacement fields in biased BLG superconductors~\cite{BLGsuperconductivity,ZhangSC2023} and exciton condensates~\cite{PhysRevLett.110.146803}. In both cases, the conventional stiffness is independent of the magnitude of the order parameter. However, the geometric stiffness depends on the magnitude of the order parameter squared, which must be attained self-consistently. Therefore, we first perform a mean-field analysis of superconductivity and exciton condensation in dual-gated bilayer graphene in the presence of a mass term $m$ due to broken $\mathcal{C}_2$ symmetry. The mass term arises due to the displacement field between the layers. We assume a momentum-independent order parameter at the Fermi surface in both cases. Inversion asymmetry, characterized by $m$, enhances the superconducting and excitonic gap, increasing the critical temperature $T_c$. We find that for small changes of the mass term, $ m \sim 0 - 50$ meV, the self-consistent value of the superconducting gap increases by several orders of magnitude, even though, the associated critical temperature, $T_c$, remains quite small. The enhancement with $m$ of the exciton gap is less pronounced, but the $T_c$ can be larger. Furthermore, The exciton gap is maximized for a density-dependent optimal value of the mass $m(n)$ suggesting that there is an ideal combination of displacement field and total density.

The geometric superfluid density for the two cases exhibits very different behaviors. The conventional stiffness, which is proportional to the density and inversely proportional to the effective mass, is similar for both cases. In the case of superconductors, the geometric superfluid density enhancement with $m$ tracks the enhancement of the superconducting but remains negligible compared to the conventional superfluid density. This is due to the small values of the superconducting gap. For the exciton condensate, the geometric contribution is of the same order as the conventional superfluid density. As a result, the geometric superfluid stiffness for the exciton condensate exhibits a rich phenomenology, the most striking of which is a density-dependent maximum value as a function of the mass term $m$. We analyze the Berezinskii-Kosterlitz-Thouless (BKT) transition temperature ($T_\mathrm{BKT}$) for the exciton condensate~\cite{KTtransition}. The additional geometric superfluid density increases the ratio $T_\mathrm{BKT}/T_c$.

The paper is organized as follows: Section I reviews chiral two-dimensional electron systems' geometric and topological properties. In section II, we solve the self-consistent gap equation for the superconductor, and the exciton condensates as a function of the density $n$ and the mass $m$. In section III, we calculate the superfluid density and discuss the enhancement of the geometric superfluid stiffness due to the $m$ for both cases. Finally, section IV discusses the influence of the mass term on the BKT transition in exciton condensates. We conclude by discussing the relevance of our results to experiments in superconducting and exciton condensates in biased BLG.    


\section{Chiral 2DEGS: Geometric properties}

The electronic properties of single and multi-layer graphene 2D crystals are described by the chiral two-dimensional electron gas (C2DEG) Hamiltonian. This class of $\kk\cdot\pp$ Hamiltonians is defined for the electron envelope wavefunction momenta $\kk$ near the Dirac points $K$($K'$) denoted by $\eta=\pm$. The chiral 2DEG Hamiltonian: $\mathcal{H}_{0,\eta}= \sum_{{\bf k},\eta} {\bf c}^{\dagger}_{{\bf k},\eta} \hat{\mathcal{H}}_{0,\eta} {\bf c}_{{\bf k},\eta}$ with chirality index $J$,
\begin{equation}
\label{C2DEGHam}
\hat{\mathcal{H}}_{0,\eta}= \zeta_J k^J \bigg( \cos(J \varphi_{\bf k}) \hat{\sigma}_x + \eta \sin(J \varphi_{\bf k}) \hat{\sigma}_y\bigg)+ m \hat{\sigma}_z,
\end{equation}
captures a chirality-dependent electronic dispersion. In Eq.~(\ref{C2DEGHam}), $\hat{\sigma}_{i}$ are the Pauli matrices defined in the sublattice space,
$\kk=(k_x,k_y)$ is the two dimensional momentum ($k=|\kk|$),
$\varphi_{\mathbf{k}} = \tan^{-1}(k_y/k_x)$ and $\zeta_J$ denotes a constant in units of eV$/$nm$^{J}$, $\zeta_1=\hbar v$ for graphene and $\zeta_2=(\hbar v)^2/\gamma_1$ ($\gamma_1 \sim 0.4$ eV) for bilayer graphene systems, with $v \approx 1 \times 10^6$ m/s. The Hamiltonian acts on the two-component spinor ${\bf c}^{\dagger}_{\mathbf{k},\eta}= (c^{\dagger}_{\mathbf{k},A,\eta}, c^{\dagger}_{\mathbf{k},B,\eta}$), with $c^{\dagger}_{\mathbf{k},A(B),\eta} (c_{\mathbf{k},A(B),\eta})$ denoting the creation (annihilation) fermionic operator for a given sub-lattice $A(B)$, valley $\eta$.
In the remainder, we use the C2DEG Hamiltonian to describe bilayer graphene.
The effect of trigonal warping is negligible for the density range
$n_{2D} = 10^{11} - 10^{12} {\rm cm}^{-2}$ for both the superconducting and exciton condensates, and therefore is not included.

The mass term $m$ breaks $\mathcal{C}_2$ symmetry, however, the system still retains particle-hole and time-reversal symmetry, expressed as $ \sigma_{y} \hat{\mathcal{H}}_{0,\eta} \sigma_y = - \hat{\mathcal{H}}_{0,-\eta}^{T}$ and $ \hat{\mathcal{H}}_{0,\eta} =  \hat{\mathcal{H}}_{0,-\eta}^{T} $ respectively. The energy dispersion for the inversion asymmetric chiral 2DEG is $\epsilon_{\kvec,J} = \pm  (\zeta_J^2 k^{2J} + m^2)^{1/2}$. For biased BLG, $J=2$ the density of states is enhanced for $ \epsilon \approx m$,
\be 
D_{J=2} (\varepsilon) = \frac{N(0) \epsilon}{\sqrt{\varepsilon^2 - m^2}} \Theta (\epsilon^2 - m^2),
\ee
where $N(0) = \gamma_1/(4 \pi (\hbar v)^2)$ and $\Theta$ is the Heaviside function. This divergent behavior at $ \epsilon \approx m $ is critical for the superconducting and exciton gap in biased BLG, as discussed in the next section.

\begin{figure}
\begin{center}
    \includegraphics[width=0.49\textwidth]{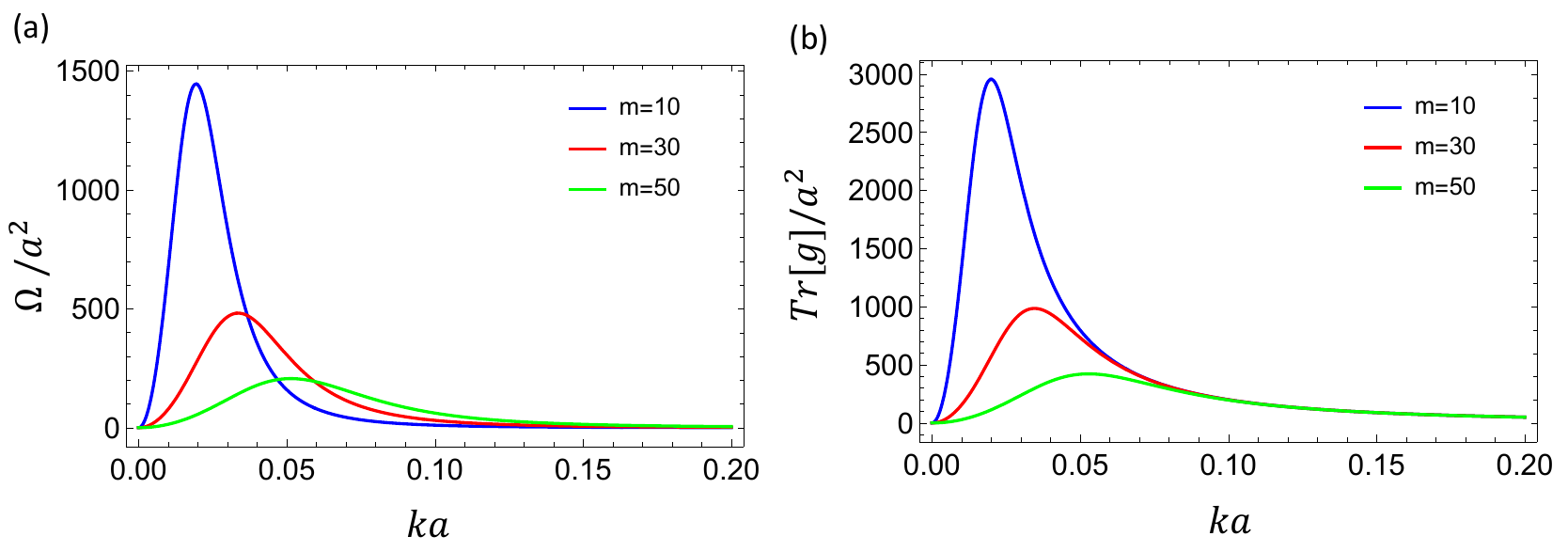}  
	\caption{a) Berry curvature $\Omega(\kk)$ and b) trace of the $g_{\mu \nu}(\kk)$ in units of $a^2$ as a function of $ka$ for different values of the displacement field, $a \sim 0.246$ nm is graphene's lattice constant.}
\label{BCandQMplot}
 \end{center}
\end{figure}

The quantum geometry of a band can be described by a gauge-invariant complex tensor $R_{\mu \nu} ({\bf k})$, called the quantum geometric tensor,~\cite{provost1980riemannian}
\be 
R^{\alpha}_{\mu \nu} ({\bf k}) = 2 {\rm Tr} \big[ P_{\alpha}({\bf k}) \partial_{\mu} P_{\alpha}({\bf k})  \partial_{\nu} P_{\alpha}({\bf k})  \big]
\ee 
where $ P_{\alpha}({\bf k}) = | u_{\alpha}(\kk) \rangle \langle u_{\alpha}(\kk)|$ denotes the projection operator for the $\alpha^{th}$-band and $\partial_{\mu} = \partial/(\partial k_{\mu})$, $\mu,\nu = k_x,k_y $ denote directions in momentum space. The real part of the quantum geometric tensor, the quantum metric, denoted by $g^{\alpha}_{\mu \nu} ({\bf k}) = \mathcal{R}e [R^{\alpha}_{\mu \nu} ({\bf k})]$ provides a notion of a quantum distance between projected states in the Hilbert space. The imaginary part $ \mathcal{I}m [R^{\alpha}_{\mu \nu} ({\bf k})]= \Omega^{\alpha}_{\mu \nu}$ is the well-known Berry curvature.

For any two-band model, $ \mathcal{H} =  \sigma \cdot  \bf n({\bf k})$, the quantum geometric tensor $R^{\alpha}_{\mu \nu } ({\bf k})$ can be expressed as,
 \be 
 R^{\alpha}_{\mu \nu} =\frac{1}{2} \partial_{\mu} \hat{{\bf n}} \cdot \partial_{\nu} \hat{{\bf n}}+  \imath \alpha \frac{1}{2} \hat{{\bf n}} \cdot  \partial_{\mu} \hat{{\bf n}} \times \partial_{\nu} \hat{{\bf n}},
 \ee
where $\alpha = \pm $ denotes the particle/hole bands, and $\hat{{\bf n}}  = {\bf n}/|{\bf n}|$. The quantum metric for the massive chiral 2DEG model is valley- and spin-independent
 \be 
g^{\pm}_{\mu \nu} (\kk) = \frac{\zeta_J^2  J^2 k^{2(J-1)} }{2(\zeta_J^2 k^{2J} + m^2)^2} \big(m^2\delta_{\mu \nu} + \zeta_J^2 k^{2J-2} (k^2 \delta_{\mu \nu} - k_{\mu} k_{\nu} )\big).
\ee
The Berry curvature $\Omega^{\eta \alpha} (\kvec)$ for a chiral 2DEG is valley-dependent,
\begin{equation}
\Omega^{\pm}_{\eta} (\kvec) = \mp \eta  \frac{\zeta_J^2 J^2}{2}  \frac{ m  k^{2(J-1)}}{(\zeta_J^2 k^{2J} + m^2)^{3/2}}.
\end{equation}

The Berry curvature $\Omega(k)$ and $Tr[g(k)]$ for biased-bilayer graphene ($J=2$) are plotted in Fig.~\ref{BCandQMplot} (a) and (b) for different values of $m$. Both exhibit a mass-dependent maximum at a non-zero wavevector. The position of the maximum in $\Omega(k)$ and $Tr[g(k)]$ scales as $0.84 (0.86) \sqrt{m/\zeta_2}$, respectively. This feature is important to the density and mass-dependent behavior of the geometric superfluid stiffness. Additionally, since the positive-definiteness of the quantum geometry tensor requires that the real part and imaginary part satisfy the relation $Tr[g ({\bf k})] \geq |\Omega^{\eta \sigma} (\kvec)|$, these functions track each other.

\section{Mean field theory for superconductivity and exciton condensation}

\subsection{Superconductivity in biased BLG}
Initial experiments on superconductivity in BLG indicate an unconventional pairing mechanism, possibly resulting in a sign-changing order parameter on the Fermi surface,~\cite{BLGsuperconductivity,ZhangSC2023,PhysRevB.105.L100503}. However, a k-independent order parameter on the Fermi surface has also been proposed as a viable candidate~\cite{ChubukovLevitovSC}.

The main goal of the present work is to elucidate the role of the quantum metric in determining the 
superfluid stiffness of superconducting and exciton-condensate states in BLG when a band gap is present.
The symmetry of the pairing has only minor quantitative effects on the superfluid stiffness, and, 
as discussed in Section IV, the s-wave pairing case provides an upper bound for the geometric contribution
to the superfluid stiffness. 
For these reasons, in treating the superconducting case, we limit ourselves to the case of s-wave pairing
%
%
%
for which the 
mean-field Bogoliubov-de Gennes Hamiltonian is 
$H_{BdG} = \sum_{\kk} \psi_{\kk}^{\dagger} \hat{H}_{BdG} \psi_{\kk}$,
with $\psi^{\dagger}_{\kk} = ({\bf c}^{\dagger}_{\kk,\upspin,+}, \bf{c}_{-\kk,\dspin,-},)$
the four-component Nambu spinor basis and
\be 
\hat{H}_{BdG} = \left( \begin{array}{cc}
\hat{\mathcal{H}}_{0,+}(\kk) & \Delta_{\kk}\\
\Delta^{\dagger}_{\kk}  & - \hat{\mathcal{H}}_{0,-}^{T} (-\kk)
\end{array} \right).
\ee
The order parameter $\Delta_\kk$ is determined from the self-consistent gap equation, 
\be 
\label{selfconsgapeq}
\Delta_{\kk}  =  \int \frac{d^2{\kk}'}{(2 \pi)^2} V_{\kk,\kk'} \Gamma_{sc} (\kvec,\kk')  \frac{\Delta_{\kk'}}{2 E_{\kk'}},
\ee
where $ E^2(\kk) = \xi_{\kk}^2 + \Delta_{\kk}^2$ with $\xi_{\kk} = (\epsilon_{\kk,J} - \mu)$, and $ \mu $ denotes the Fermi energy. We assume $ \mu >0$ (the results for $\mu <0$ can be easily attained from the $\mu>0$ results considering the particle-hole symmetry of the bilayer graphene Hamiltonian). $\Gamma_{sc}(\kk,\kk')$ denotes the superconducting form factor,
\be
\Gamma_{sc} (\kvec,\kk') =
 \frac{1}{2} \big( 1+\cos \theta_{\kvec} \cos \theta_{\kk'} \big),
\ee
where $\cos(\theta_{\kvec,J}) = m/|\varepsilon_{\kvec,J}|$.


To calculate the enhancement of the quantum metric contribution due to the superfluid weight, 
the detailed values of the parameters $V_0$ and $\omega_c$ are not important as long as they return the correct value
of $T_c$, i.e., of the amplitude of the superconducting order parameter $\Delta$. We estimate $\omega_c \sim \hbar v_s k_{F}$, 
where $v_{s} \sim 1 \times 10^6$ cm/s is the phonon sound velocity in graphene \cite{PhysRevB.105.L100503} and determine the value of $V_{0}$
which corresponds to the experimentally measured value of $T_c$ \cite{BLGsuperconductivity}. Considering that for $n\sim 6\times 10^{11}{\rm cm}^{-2}$, 
and $m=50$~meV, the experimentally measured $T_c$ is $\sim 30$~mK, and $\omega_c \sim 1$ meV we find 
$V_0=908\;{\rm meV\cdot nm^2}$. 
This value is similar in magnitude to the phonon-mediated attractive interaction estimate in Ref.~\cite{PhysRevB.105.L100503}.


\begin{figure}
\begin{center}
    \includegraphics[width=0.49\textwidth]{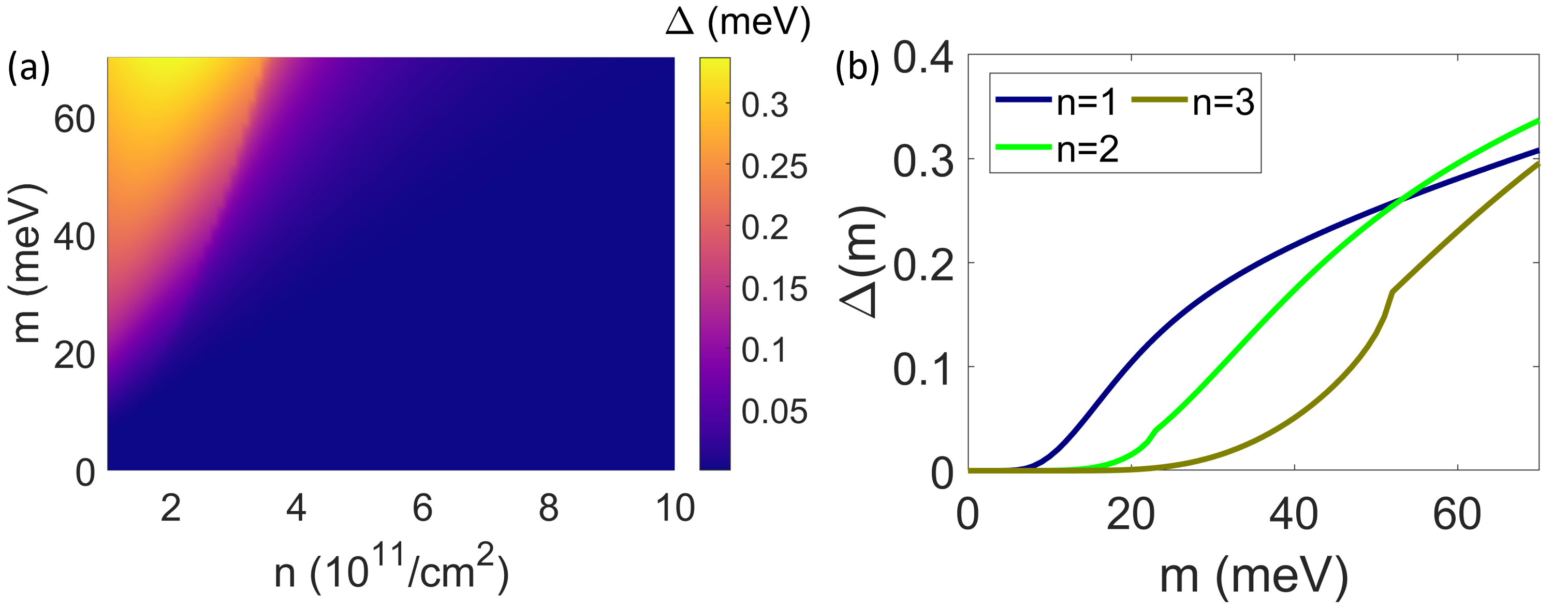}  
	\caption{The magnitude of the superconducting order parameter $\Delta$ as a function of the mass gap and two-dimensional charge density $n$ for bilayer graphene. Here $V_{0}=908~$ meV$\cdot$ nm$^2$ and $\omega_c = 1$ meV. The unit of $n$ is $10^{11}$cm$^{-2}$ if not specified specially. }
\label{SCgapplot}
 \end{center}
\end{figure}

The order parameter grows exponentially with $m$ for $m \sim 0 -50$ meV, as shown in Fig.~\ref{SCgapplot} (a) and (b). It then increases linearly for higher values of the mass term. This enhancement results from the large density of states at low values of $n$. For $m=0$, the density of states in BLG is constant. Therefore, $ \Delta $ has no dependence on the carrier density $n$. However, when $m \neq 0$ at lower values of densities, there is a large density of states, resulting in an enhancement of the order parameter, as indicated in different line cuts in Fig.~\ref{SCgapplot} (b). 
The full phase diagram as a function of the mass $m$ and the two-dimensional density $n$ is plotted in Fig.~\ref{SCgapplot} (a). Next, we study the exciton gap in bilayer exciton condensates, which shows different behavior from the superconducting gap due to the long-range nature of the attractive interaction.

 \begin{figure}
 \begin{center}
    \includegraphics[width=1.0\linewidth]{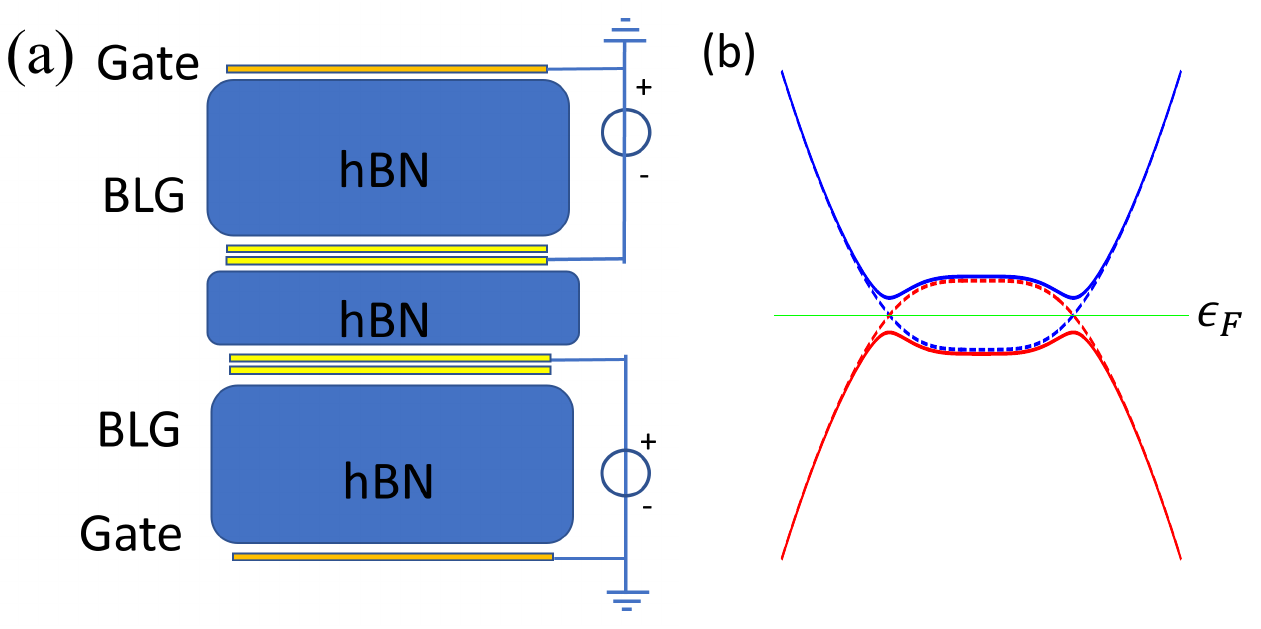}	
	\caption{(a) Schematic of dual-gated bilayer graphene device for exciton condensate. (b) The quasiparticle bands dispersion in biased bilayer graphene exciton condensate. Dashed lines indicate the non-interacting electron-hole bands. The Fermi energy is indicated in green.}
	\label{F1device}
	\end{center}
\end{figure}

\subsection{Exciton condensation in biased double BLG}
Since indirect excitons in spatially separated electron-hole bilayer are protected from recombination~\cite{Eisenstein2004}, these systems are ideal platforms for exciton (electron-hole bound pair) condensation~\cite{PhysRev.126.1691}. To realize BLG excitons condensates, we therefore consider two BLG systems separated by a dielectric of thickness $d$ (see Fig.~\ref{F1device})~\cite{PhysRevLett.110.146803,PhysRevB.78.121401,PhysRevB.77.233405,PhysRevB.78.121401,PhysRevB.77.233405,lutchyn2010d}. The densities of electrons and holes in the upper or lower bilayer graphene can be tuned by an electric field applied perpendicular to the combined hetero-structure, as indicated in Fig.~\ref{F1device}~(a). The layers are gated separately with the gate potential ($V_{g},-V_{g}$) in the top and bottom layers. The gate potential is adjusted for a p-type Fermi surface(FS) in one layer nested with the n-type FS in the other layer. The setup in the figure allows independent control of the doping and layer potential.
Electrons in the top layer pair up with holes in the bottom layer, forming excitons, and giving rise to a gapped spectrum 
as shown in Fig.~\ref{F1device}~(b).  
Particle-hole (PH) symmetry in exciton condensates plays the same role as time-reversal symmetry in superconductors~\cite{PhysRevLett.101.256406}. 2D crystals generally satisfy perfect particle-hole symmetry~\cite{PhysRevB.78.121401,PhysRevB.77.233405}, and therefore they are attractive candidates to realize double-layer exciton condensates~\cite{PhysRevB.78.121401,PhysRevB.77.233405}.

The mean-field exciton parameter, here denoted by $\Delta^{\perp}$, is calculated from a self-consistent gap equation similar to  Eq.~\ref{selfconsgapeq} (see the Appendix A for details) assuming for the inter-layer interaction $V_{D} (q) = 2 \pi e^2/(\epsilon q) e^{-qd}$. The self-consistent gap equation for $\Delta^{\perp}$ can be written in the form (see also Appendix A):
\be 
\label{eq:excitongap}
1 = \kappa \int_{-\pi/2}^{\pi/2} d \varphi \int^{2 \cos \varphi}_{0} d \bar{q} \frac{e^{-\bar{q} k_F d} \cos \theta_{\kk - \qq} \cos \theta_\kk}{\sqrt{(\bar{\Delta}^{z}_{\kk - \qq})^2 + (\bar{\Delta}^{\perp})^2}} \bigg|_{|\kk| = k_F},
\ee
where $\kappa = e^2 \gamma_1/(4 \pi \epsilon (\hbar v)^2 k_F)$ is the coupling constant, and the bar denotes that all energies are measured relative to the Fermi energy $\epsilon_F$, and the exciton gap $\bar{\Delta}^{\perp}$ and chirality factors are all evaluated at $|\kk| = k_F$. The $k_F^{-1}$ dependence of the coupling constant and exponential decay as a function of $k_F$ results from the Coulomb interaction. In the calculation for $\Delta^{\perp}$ we neglect the effective mass renormalization of $\Delta^z$ in Eq.~\ref{eq:GapeqEX} due to intra-layer interaction, as it has been shown to have a marginal effect on the value of the exciton gap in BLG~\cite{PhysRevLett.111.086804}. Before we discuss the results of the exciton gap, we justify our use of unscreened Coulomb interactions in   Eq.~\ref{eq:excitongap}. 

The calculation of the strength of the electron-hole Coulomb interaction for electron-hole bilayers, 
taking correctly into account screening effects, is subtle. This is due to the difficulty of self-consistently accounting for the effect of the
formation of the exciton condensate on the screening of the interlayer Coulomb interactions.
This issue has been extensively studied in the literature. Using a Thomas-Fermi {\em static} screening approximation 
Refs.~\cite{PhysRevB.78.241401}, and ~\cite{Kharitonov_2010},
concluded that the exciton condensate's order parameter $\Delta$
might be vanishingly small. However, follow-up works
\cite{bistritzer2008commentelectronscreeningexcitonic,PhysRevB.85.195136,PhysRevB.88.235402,PhysRevResearch.5.043176,PhysRevB.95.045416,PhysRevLett.133.056501},that more carefully treated the competition between the screening and opening of a band-gap due to the 
establishment of an exciton condensate,returned much higher values of $\Delta$. While electron density reduces the strength of the interaction, at the same time,
the effect of opening a gap due to the formation of the exciton condensate strongly suppresses the screening.
Different results are obtained depending on how the two effects are considered.
There are three possible approximations to treat the screening of the electron-hole interaction:
(i)   Unscreened (US) approximation in which no screening effect is included;
(ii)  Normal screening (NS), in which screening is included using the random phase approximation in the normal state;
(iii) Superfluid state (SS) screening, in which screening is included using the random phase approximation
in the superfluid coherent state.

By comparing the results obtained using the three different approximations to highly accurate diffusion quantum Monte Carlo 
(DQMC) results Neilson {\it et. al.} in Ref.~\cite{PhysRevB.89.060502} showed that
the SS approximation is the most accurate. In addition, they showed that when the interaction parameter $r_s$ is larger
than 3, the SS and US approximations return almost undistinguishable results. 
For gapless bilayer graphene we have that $r_s>3$ corresponds to densities $n<7\times 10^{12} {\rm cm}^{-2}$. 
The bands flatten for a nonzero band-gap $m$; therefore, for fixed density, $r_s$ increases. As a consequence, 
as long as  $n<7\times 10^{12} {\rm cm}^{-2}$ the US approximation should be used to estimate the size of the
exciton condensate order parameter $\Delta^\perp$, for all values of the gap $m$.
In the remainder, we consider densities only up to $10^{12} {\rm cm}^{-2}$,
well within the range of validity of the unscreened approximation.
To take into account the screening effects of the nearby gates we use a fairly large value of the dielectric constant setting
$\epsilon=10$.

 \begin{figure}
 \begin{center}
    \includegraphics[width=0.49\textwidth]{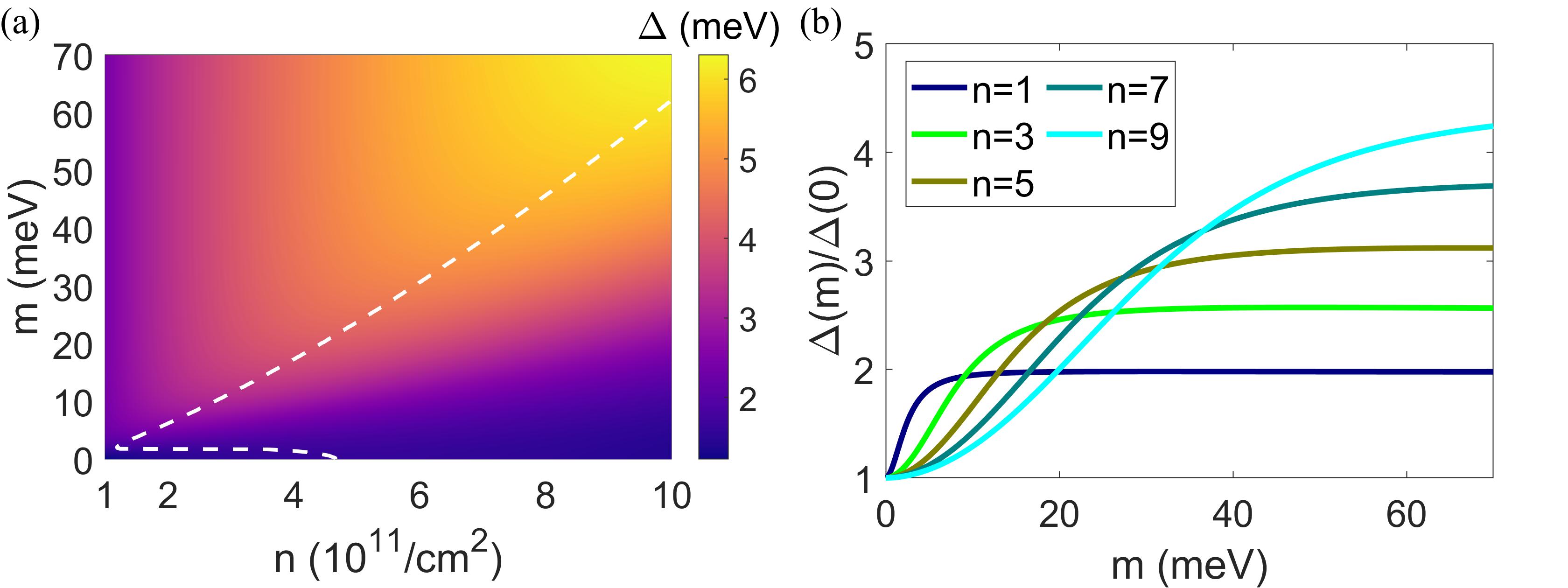}  
	\caption{The magnitude of the exciton order parameter $\Delta$ as a function of the mass gap and two-dimensional charge density $n$ for biased bilayer graphene. The inset in (b) shows the $\Delta(m)$ at lower values of $m$. The white dotted line in (a) indicates the optimal density which corresponds to the maximum exciton gap for each $m$.}
\label{fig:DtJ2EC}
\end{center}
\end{figure}
 
The results for the exciton gap with Coulomb inter-layer interactions are plotted in Fig.~\ref{fig:DtJ2EC} (a) as a function of $m$ and $n$ for double bilayer graphene.  We take the Fermi energy $E_F \sim 3 - 35 $ meV at $m=0$ which corresponds to the densities $n=  0.1-1 \times 10^{12}$ cm$^{-2}$ and $d=1 nm$. At $m=0$, the exciton gap is expected to decrease as a function of the density $n$ (not visible due to the scale in Fig.~\ref{fig:DtJ2EC} (a). 

Fig.~\ref{fig:DtJ2EC} (a) indicates that the exciton gap is maximized for an optimal value of $n$ and $m$, indicated by the white dotted line. This behavior can be understood by studying the low and high $n$ limits for $m$ in Eq.~\ref{eq:excitongap}. At low densities, electrons reside in flat-band regions with access to a large density of states, resulting in a sudden increase in the exciton gap. At large densities, the exponential term in Eq.~\ref{eq:excitongap} dominates, thereby reducing the exciton gap sharply. At intermediate values of the density $n$, these trends conspire to produce a density-dependent local maximum value of the exciton gap. This analysis indicates an optimal value of displacement fields to search for exciton condensates in biased bilayer graphene. 
 
The exciton gap enhancement as a function of $m$ is shown in Fig.~\ref{fig:DtJ2EC} (b). Even though the exciton gap is enhanced as a function of $m$, this enhancement is less pronounced than that of the superconducting gap. The exciton gap exhibits a steady enhancement as a function of the mass $m$ which saturates to a density-dependent value of the exciton gap enhancement. There is a larger enhancement at higher densities, with a higher gradient of enhancement associated with lower densities. These differences are due to the long-range nature of the Coulomb interaction, which results in a density-dependent coupling parameter.

\section{Superfluid density}

While the paring interaction in the superconductor and exciton condensate corresponds to physically distinct processes, the mathematical structure of the mean-field Hamiltonian is similar, allowing for a unified description of the superfluid properties. 
 \begin{figure}
 \begin{center}
    \includegraphics[width=1.0\linewidth]{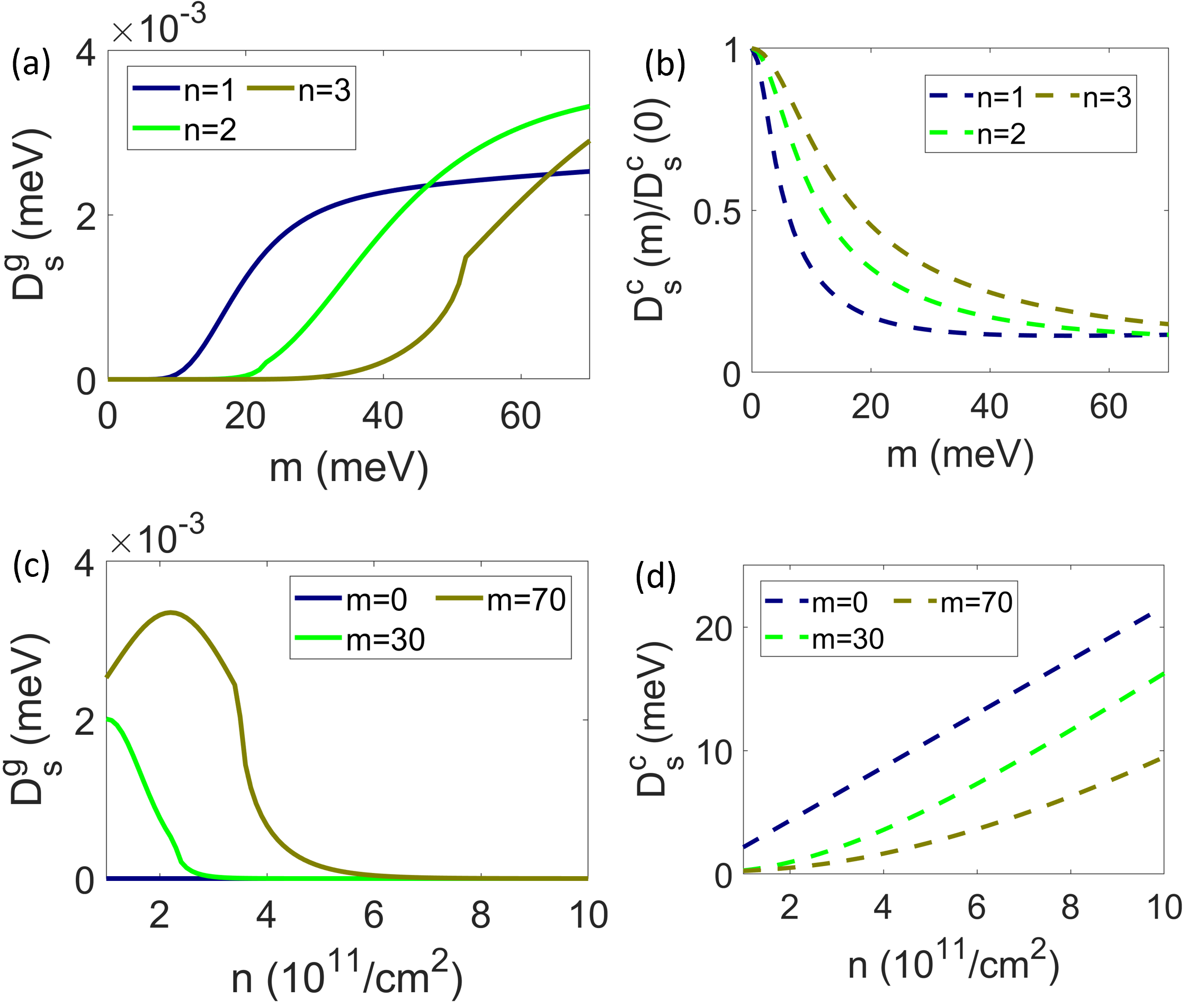}	
	\caption{The superfluid stiffness for bilayer graphene superconductor. (a) \& (c) plot the geometric stiffness as a function of $m$ for different density values $n$ and as a function of $n$ at different $m$, respectively. (b) The conventional superfluid density as a function of $m$ for fixed $n$ and (d) as a function of $n$ for fixed $m$. The solid line labels the geometric term in all graphs, while the dashed line is the conventional superfluid density. The unit of energy is meV.}
	\label{fig:DsJ1J2cut}
	\end{center}
\end{figure}
The superfluid density of the condensate characterizes its ability to carry a supercurrent. For a two-band model, taking the Fermi energy to lie within the $\alpha^{th}$ band gives two contributions to the superfluid density for the $\alpha^{th}$-band $D_{\mu \nu} = D^{conv}_{\mu \nu} +D^{geo}_{\mu \nu}$,~\cite{PhysRevB.95.024515} where,
\be
\label{eq:Dsconv}
D^{conv}_{\mu \nu} =  \frac{g}{L^2} \sum_{\kvec} \bigg( 1 - \frac{\xi_{\alpha,\kk}}{E_{\alpha,\kk}} \tanh \big(\frac{\beta E_{\alpha,\kk}}{2} \big)\bigg)  \partial_{\mu } \partial_{\nu} \epsilon_{\kk,\alpha},
\ee
is the conventional contribution, and
\be
\label{eq:Dsgeo}
D_{\mu \nu}^{geo}= \frac{g}{L^2} \sum_{\kk,\alpha = \pm}  \frac{2|\Delta_{\kk}|^2}{\alpha E_{\kk,\alpha}}\tanh \big(\frac{\beta E_{\kk}}{2} \big) g_{\mu \nu} (\kk),
\ee
is the geometric contribution. In both Eq.~\ref{eq:Dsconv} and ~\ref{eq:Dsgeo}, $E_{\kk,\alpha}$ denotes the quasiparticle dispersion of the BLG superconducting or exciton condensate, and $\beta=1/(k_{\beta} T)$. At $T=0$, $D_{\mu \nu}^{conv} = n/m^{\star} \delta_{\mu \nu}$, where 
$m^{\star} =\hbar^{2} (\partial^2 \epsilon_{\kk}/ \partial k^2)^{-1}$
is the effective mass. This result is independent of the symmetry of the order parameter. The $s$-wave order parameter maximizes the geometric superfluid stiffness, as any non $s$-wave symmetry reduces phase space in the integral over momentum space in Eq.~\ref{eq:Dsgeo}. Therefore, our results for the s-wave symmetry provide an upper bound for the geometric superfluid density associated with the superconductor and exciton condensates in biased bilayer graphene.

Using azimuthal symmetry we can write $D_{xy}^{geo} =0$ and $D_{xx}^{geo} = D_{yy}^{geo} = D^{geo}$. For $m=0 $, the total superfluid density can be expressed as,
\be 
D_s =  \frac{g J}{2 \pi} \sqrt{\mu^2 + \Delta^2} + \frac{g J\Delta^2 }{2\pi \mu}  \log  \bigg( \frac{\mu + \sqrt{\Delta^2 + \mu^2} }{\mu - \sqrt{\Delta^2 + \mu^2} } \bigg). 
\ee
For weak coupling, $\mu = \epsilon_F$ and $\Delta \ll \epsilon_F$, resulting in a comparatively minor geometric superfluid stiffness. For $m>0$, the superfluid density is calculated numerically using Eqs.~\ref{eq:Dsconv} and \ref{eq:Dsgeo}. 

The geometric and conventional superfluid densities as functions of $m$ for the superconducting case 
are shown in Fig.~\ref{fig:DsJ1J2cut}. As in the case $m=0$, the geometric superfluid stiffness is negligible $D_s^{geo}/D_s^{conv} \sim 10^{-4}$ for $m$ up to $\sim 50 $ meV due to the small superconducting gap. The geometric superfluid stiffness is plotted as a function of $m$ in Fig.~\ref{fig:DsJ1J2cut} (a); it follows the same behavior as the enhancement of the superconducting gap. As expected, the conventional superfluid stiffness decreases as a function of $m$ as indicated in Fig.~\ref{fig:DsJ1J2cut} (b), due to an increase in the effective mass $m^{\star}$. 

 \begin{figure}
 \begin{center}
    \includegraphics[width=1.0\linewidth]{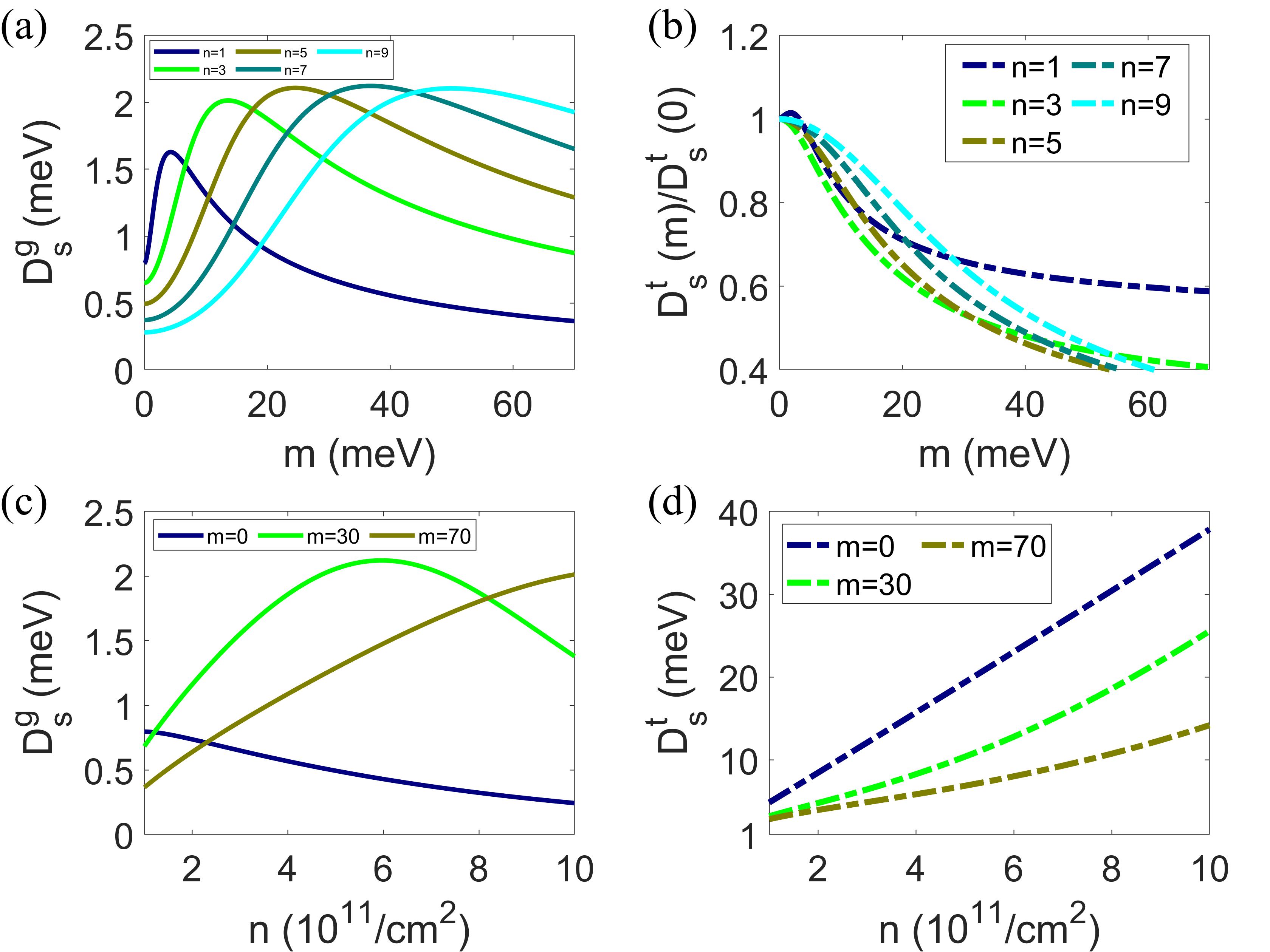}	
	\caption{The superfluid stiffness for the exciton condensate. (a) \& (c) plot the geometric term as function of $m$ for different values of density $n$ and as a function of $n$ at different $m$, respectively. (b) The ratio of the superfluid density as a function of $m$ for fixed $n$. (d) Total superfluid density as a function of $n$ for fixed $m$.  In all graphs, the solid line labels the geometric term, while the dashed line is the total superfluid density. The unit of energy is meV.}
	\label{DsJ2cutEC}
	\end{center}
\end{figure}

In contrast, the geometric superfluid stiffness for the exciton condensate exhibits a much richer phenomenology as indicated in Fig.~\ref{DsJ2cutEC} (a-d). Due to the large value of the exciton order parameter, the geometric superfluid stiffness 
is non-negligible, contributing up to $\sim 20\%$ of the total stiffness, and 
has a density-dependent maximum as a function of $m$, as seen in Fig.~\ref{DsJ2cutEC} (a).
The geometric superfluid stiffness acquires a $n$ and $m$ dependent maximum, as shown in Fig.~\ref{DsJ2cutEC} (c). 
The scaling with respect to $n$, and $m$, of the geometric contribution to $D_s$ 
is determined by the trend of the Berry curvature and quantum metric trace, 
see Fig.~\ref{BCandQMplot}, and the trend of the exciton gap, 
see Fig.~\ref{fig:DtJ2EC}.
The conventional term still determines the overall trend of the total superfluid density, except at very low densities
 $n \lesssim 1 \times 10^{11}$ cm$^{-2}$, for which there is a slight enhancement due to the geometric contribution,
see Fig.~\ref{DsJ2cutEC} (b).

\section{Berezinskii-Kosterlitz-Thouless transition}

In superconductors and superfluids, the BKT phase transition separates the superfluid and resistive states and is associated with the binding-unbinding of vortices.~\cite{KTtransition} The critical temperature of the BKT phase transition $T_{\mathrm{BKT}}$ is determined from the relation $k_B T_{\mathrm{BKT}}=\pi D_s(\Delta(T_{\mathrm{BKT}}), T_{\mathrm{BKT}})/8$, where $D_s$  is the total superfluid density. A direct consequence of a geometric contribution to $D_s$ is an increase of $T_{\mathrm{BKT}}$. In the following, we only present $T_{\mathrm{BKT}}$ for the exciton condensate. The results for the superconductors, which exhibit negligible geometric superfluid stiffness, follow the standard relations.~\cite{KTtransition}

 \begin{figure}[tb]
 \begin{center}
    \includegraphics[width=1.0\linewidth]{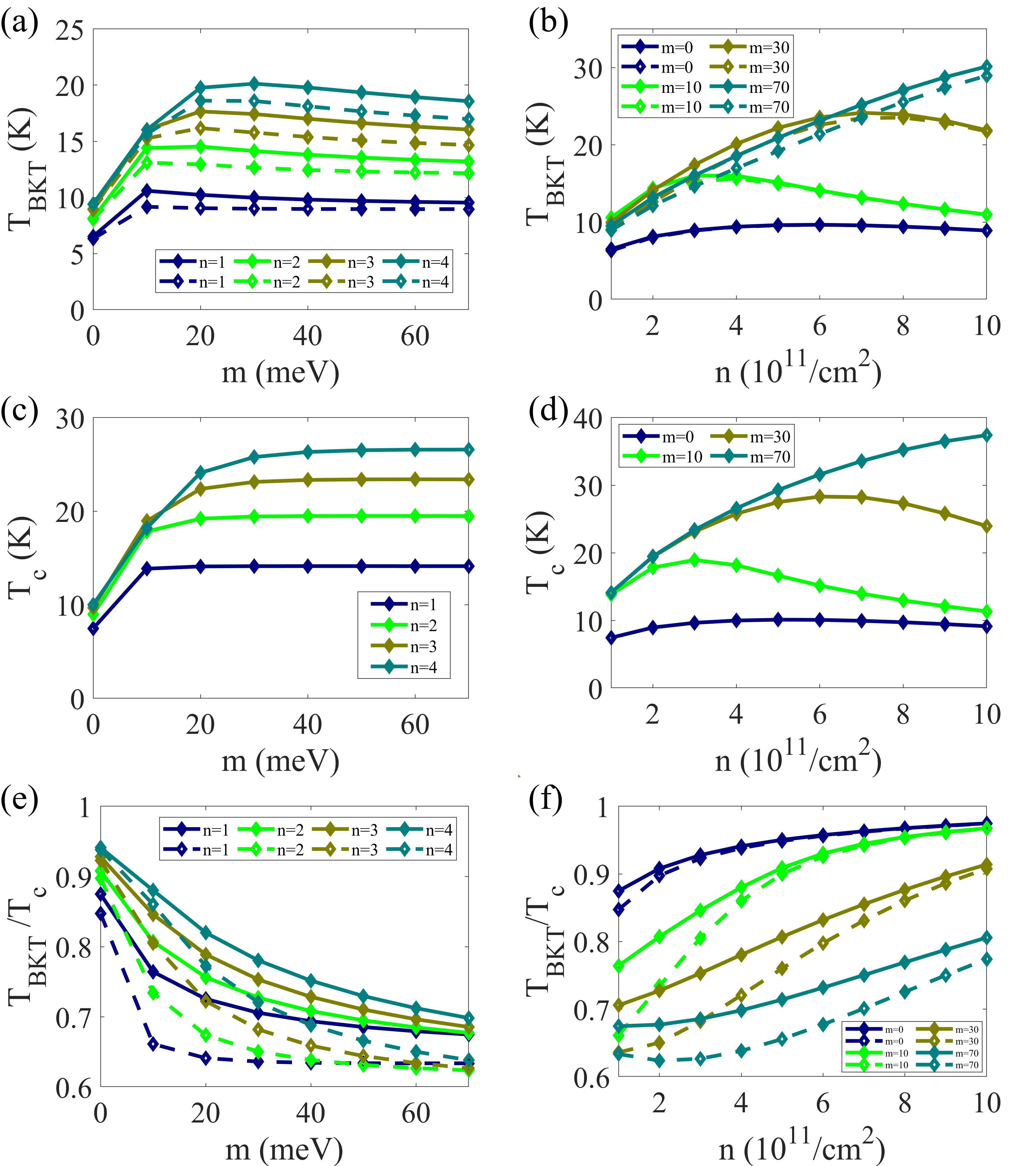}	
	\caption{The exciton condensate BKT transition temperature $T_{\mathrm{BKT}}$, (a) as a function of $m$ for different values of density $n$ and (b) as a function of $n$ at different $m$. In both (a) and (b), the dashed lines correspond to $T_{\mathrm{BKT}}$ calculated using only the conventional superfluid weight, while the solid lines show the $T_{\mathrm{BKT}}$ using the total superfluid weight. Note that in (b), for $m<70$, the differences between the solid lines and their corresponding dashed lines are very small. The exciton condensate's critical temperature $T_c$ (c) as a function of $m$ for different $n$ and (d) as a function of $n$ at different $m$. The ratio $T_{\mathrm{BKT}}/T_c$, (e) as a function of $m$ for different $n$ and (f) as a function of $n$ at different $m$. As in panels (a) and (b), also in panels (e) and (f) the dashed and solid lines show the results with, and without, the geometric contribution, respectively.}
	\label{J2BKTEC}
	\end{center}
\end{figure}

Fig.~\ref{J2BKTEC} (a-f) summarize our findings for $T_{\mathrm{BKT}}$ and critical temperatures $T_c$ for the exciton condensate as a function of the density $n$ and mass $m$. The dotted lines show $T_{\mathrm{BKT}}$ calculated with just the conventional term, while the solid lines show the value of $T_{\mathrm{BKT}}$ obtained taking into account the geometric contribution. As expected, adding the geometric superfluid density slightly enhances $T_{\mathrm{BKT}}$. The enhancement of the ratio $T_{\mathrm{BKT}}/T_c$ is more pronounced at lower densities and higher masses as indicated in Figs.~\ref{J2BKTEC} (e) \& (f).

\section{Discussion and Outlook}

Inversion asymmetry due to a mass term $m$ enhances the superconducting and exciton gap in biased-BLG and double-BLG. The superconducting gap is enhanced exponentially, prominent at low densities, due to the large density of states resulting from band flatness near the Dirac point. This is followed by a concomitant enhancement of the geometric superfluid stiffness. However, the conventional superfluid stiffness dominates in the weak coupling limit. The momentum-independent $s$-wave order parameter considered in this paper provides an upper bound for the geometric superfluid density in biased BLG. Therefore, the geometric contribution to superfluidity in bilayer graphene superconductors should be negligible.~\cite{BLGsuperconductivity}

For the exciton condensate, the band structure modification due to the mass term and the long-range nature of the Coulomb interactions produces a modest increase in the exciton gap. The exciton gap is maximized for a density-dependent optimal value of of the mass $m$ with critical temperatures $T_c \sim 1 - 10 K$. These critical temperature estimates are consistent with more sophisticated studies of exciton condensation in two-dimensional crystals~\cite{PhysRevLett.110.146803,PhysRevB.85.195136,PhysRevB.96.174504}, implying mean-field theory qualitatively captures the behavior of the exciton condensate gap in biased bilayer graphene. These larger values of the exciton condensate gap result in a non-negligible contribution to the total superfluid stiffness. 
This geometric contribution is more pronounced at lower densities and higher mass values, where flatter regions of the electronic band dispersion influence the exciton condensate. An experimental consequence of the more significant total superfluid density is an increase in the BKT transition temperatures of the exciton condensate in biased bilayer graphene.

The duality of the superconducting and exciton condensates, where time-reversal is interchanged to particle-hole 
 symmetry has been discussed in lattice models~\cite{PhysRevLett.117.045303}. Since the conventional contribution is generally independent 
 of the strength of the order parameter, it has a similar value for the superconducting and excitonic 
 condensates. On the other hand, the geometric term depends on the strength of the square of the 
 superconducting and exciton order parameters. The difference in the geometric superfluid 
 weight in these cases originates from different self-consistent superconducting and 
 excitonic gap values, which are due to the pairing mechanisms. 

\acknowledgments{The authors acknowledge helpful discussions with G. Jiang and Han Fu. Y.B. was supported by the University of Nevada, Reno VPRI startup grant PG19012. E.R. acknowledges support from DOE, grant No. DE-SC0022245. 
We acknowledge William \& Mary Research Computing for providing computational resources and/or technical support that have 
contributed to the results reported within this paper. URL: https://www.wm.edu/it/rc.  
X.H. acknowledge funding from the National Natural Science Foundation of China (Grant No. 11904143, and No. 12364022), 
the Natural Science Foundation of Guangxi Province Grant No. 2020GXNSFAA297083, GuiKe AD20297045.}	

\appendix

\section{Details of mean field theory for the exciton condensate}

Without loss of generality, we assume that the gate voltages are such that the Fermi energy lies in the conduction band for the top-layer ($\sigma=1(+))$ and in the valence band for the bottom-layer ($\sigma=2(-)$). The single particle energy dispersion of the low energy bands of hetero-structure is symmetric about the Fermi energy due to particle-hole symmetry with $e_{\mathbf{k},J,+}= - V_{g} + \varepsilon_{\kvec,J}$ and $e_{\mathbf{k},J, -}=V_{g}-\varepsilon_{\kvec,J}$ as indicated in Fig.~\ref{F1device} (b). Assuming that the exciton order parameter $\Delta$ is smaller than the applied gate potential $|V_{g}|$, the density-density interactions can be projected onto the electron band in the top layer and the hole band in the bottom layer. The interaction Hamiltonian can be expressed as, 
\begin{equation}
\mathcal{H}_{int} =  \frac{1}{2 L^2} \sum_{q,\alpha = \pm} 
\big( V_{{\bf q}} \bar{\rho}_{{\bf q},\alpha} \bar{\rho}_{-{ \bf q},\alpha} +  V^{d}_{{\bf q}} \bar{\rho}_{{\bf q},\alpha} \bar{\rho}_{-{ \bf q},-\alpha} \big),
\end{equation}
where $L^2$ denotes the area of the heterostructure, $\sigma = \pm = (t,b)$ are identified with the electron  bands in the top layer and the hole bands in the bottom layer, $V_{\mathbf{q}}^{d}$
($V_{\mathbf{q}}$) refers to the interlayer (intralayer) interaction,
and the projected density operator can be expressed as,
\begin{equation}
\bar{\rho}_{{\bf q},\alpha} = \sum_{{\bf k},\alpha}  \langle \chi_{J, \alpha} ({\bf k}+{\bf q}) | \chi_{J,\alpha} ({\bf k}) \rangle \gamma^{\dagger}_{{\bf k}+{\bf q},\alpha,\sigma} \gamma_{{\bf k},\alpha,\sigma},
\end{equation}
where $\gamma^{\dagger}_{\kvec,\alpha,\sigma} (\gamma_{\kvec,\alpha,\sigma}) $ denotes the $ \alpha^{th}$ band creation and annihilation operator at $ \kvec$, and
 $\sigma$ denotes the spin and valley degrees of freedom, which remain unaffected by the mass term $m$. The eigenfunctions $| \chi_{J,\alpha} ({\bf k}) \rangle$
\be
| \chi_{J,+} ({\bf k}) \rangle = \left( \begin{array}{c}
\cos(\theta_{\kk}/2) \\
\sin(\theta_{\kk}/2) e^{\imath J \varphi_{\kk}} \end{array} \right) 
\ee
and $  | \chi_{J,-} ({\bf k}) \rangle = \imath \sigma_y | \chi_{J,+} ({\bf k}) \rangle , $  determine the form factors in the band projected density operator.
 The form factors associated with the wavefunction overlap in the projected density determine the symmetry of the exciton order parameter $\Delta$ and the fluctuations of the exciton condensate, as we show next.

For a non-zero expectation value for the exciton order parameter one
obtains a mean-field Hamiltonian, $ \mathcal{H}_{MF}= - \sum_{\mathbf{k},\sigma,\sigma'} \gamma_{\mathbf{k},\sigma}^{\dagger}\boldsymbol{\Delta}_{\mathbf{k}}\cdot\boldsymbol{\tau}_{\sigma\sigma'} \gamma_{\mathbf{k},\sigma'},
$ where $\boldsymbol{\Delta}=(\Delta^{x},\Delta^{y},\Delta^{z})$
denote the mean-fields and $\boldsymbol{\tau}=(\tau^{x},\tau^{y},\tau^{z})$
are the $2\times2$ Pauli matrices acting in the layer
pseudospin space. The transverse components
of the pseudospin field $\boldsymbol{\Delta}_{\mathbf{k}}$
define a complex order parameter $\Delta_{\mathbf{k}}^{\perp}=\Delta_{\mathbf{k}}^{x}-i\Delta_{\mathbf{k}}^{y}$,
whose magnitude $\left|\Delta_{\mathbf{k}}^{\perp}\right|$ determines
the strength of the particle-hole condensate. The mean-fields $\boldsymbol{\Delta}_{\mathbf{k}}$
are given by the following self-consistent
equations: 
\begin{eqnarray}
\label{eq:GapeqEX}
\nonumber
\Delta_{\mathbf{k}}^{z} &=&\eta_{\mathbf{k}}+ \frac{1}{2L^2}\sum_{\mathbf{p}}\left[V_{\mathbf{k-p}}\Gamma_{ex}(\kvec,\pvec)-E_H \right] \nonumber \\
&\times& \left[1+\frac{\Delta_{\mathbf{p}}^{z}}{E_{\mathbf{p}}}
f\bigg(\frac{E_{\pvec}}{2} \bigg)\right];\label{eq:Dz} \nonumber \\
\Delta_{\mathbf{k}}^{\perp} &=& \frac{1}{2L^2}\sum_{\mathbf{p}} V_{\mathbf{k-p}}^{d} 
\Gamma_{ex}(\kvec,\pvec)\frac{\Delta_{\mathbf{p}}^{\perp}}{E_{\mathbf{p}}}
f\bigg(\frac{E_{\pvec}}{2} \bigg),\label{eq:Dperp}
\end{eqnarray}
where $\eta_{\mathbf{k}} =(e_{\kvec,J,+} - e_{\kvec,J,-})/2$, $f(x) = \tanh(\beta x) $ and $\beta = 1/(k_{\beta}T)$.
The interlayer Coulomb interaction in the direct channel, $E_H =2\pi e^{2}gd/\epsilon$
captures the layer charging energy, $g=4$ is the spin and valley degeneracy,
$\epsilon$ is
the dielectric constant of the embedding media, 
and $E_{\mathbf{k}}=\sqrt{(\Delta_{\mathbf{k}}^{z})^{2}+\left|\Delta_{\mathbf{k}}^{\perp}\right|^{2}}$. 
$\Gamma_{ex}(\kk,\kk')$ denotes the angle-dependent chirality form factor for the exicton condensate,
\begin{eqnarray}
\Gamma_{ex}(\kvec,\kk')&=&
 \frac{1}{2} \big( 1+\cos \theta_{\kvec,J} \cos \theta_{\kk',J}\nonumber\\
 &&+ \sin \theta_{\kk,J} \sin \theta_{\kk',J} \cos(\varphi_{{\kvec},{\kk'}} ) \big),
\end{eqnarray}
where $\cos(\theta_{\kvec,J}) = m/|\varepsilon_{\kvec,J}|$ and $\varphi_{\mathbf{k},\mathbf{k}'}\equiv\ J (\varphi_{\mathbf{k}}-\varphi_{\mathbf{k}'})$. In general, the chirality form factor $\Gamma_{EX}(\kk,\kk')$ results in an order parameter of the form $\Delta_{\mathbf{k}} =\left|\Delta_{\alpha}\right| e^{i \alpha \varphi_{\mathbf{k}}+i\phi}$
with chirality $\alpha=0,\pm1,\pm2,\dots$ ~\cite{PhysRevLett.111.086804} and an arbitrary global
phase $\phi=0$. Substituting $\Delta_{\mathbf{k}} = \left|\Delta_{\alpha}\right| e^{i \alpha \varphi_{\mathbf{k}}}$ in the gap equation above and integrating over $\varphi_{\kvec}$ results in only three orthogonal solutions $\alpha=0,\pm J$ for any central interactions. Solving the exciton gap equation we find that the $\alpha=0$ channel exhibits the largest gap for all values of $m$, hence, we focus on a constant value of the exciton order parameter at the Fermi surface.

\section{Superfluid density: Geometric and conventional contribution}

The Kubo formula for the superfluid density~\cite{PhysRevB.95.024515} can be expressed as, 
\bea
D_{\mu \nu} &=& \frac{g}{L^2} \sum_{\kvec, ij} \frac{n(E_j) - n(E_i) }{E_i-E_j} \bigg( \langle \psi_i| \partial_{\mu} H | \psi_j\rangle \langle \psi_j| \partial_{\nu} H | \psi_i \rangle 
\rangle \nonumber \\ &-& \langle \psi_i| \partial_{\mu} H \gamma_z| \psi_j\rangle \langle \psi_j| \gamma_z \partial_{\nu} H | \psi_i \rangle \bigg),
\eea
where $g=2(4)$ denotes the degeneracy of the superconductor (exciton condensate),  $i(j) = (\pm,\alpha)$ labels the particle-hole Bogoliubov-deGennes (BdG) bands in the $\alpha = \pm$ band, with $\psi_{i} $ and $E_{i}$ denoting the respective BdG eigenfunctions and eigenvalues, and $n(E)$ the Fermi-Dirac distribution. The first term corresponds to the diamagnetic contribution, while the second term is the paramagnetic contribution. At points of degeneracy the difference between the Fermi functions should be replaced by the derivative $dn/dE_j$.

The gauge fields in superconductors and exciton condensates correspond to physically distinct processes, namely the magnetic field for superconductors and displacement field asymmetry $\qvec = e (\vec{A}_t - \vec{A}_b)$ (where $ \vec{A}_i$  denotes the electromagnetic gauge field in the $i^{th}$-layer) for the latter. Nevertheless, the mathematical structure of the BdG mean-field Hamiltonian is the same, allowing for a unified description. In the case of the superconductor, we assume $\mu >0$. The superconducting BdG eigenstates for the $\alpha = +$ band can be expressed in the basis set $(\kk,\upspin,+; -\kk,\dspin,-)$, $ \psi^{\dagger}_{\alpha,-} (\kvec)=(u_{\alpha} ,v_{\alpha} )$ and $ \psi^{\dagger}_{\alpha,-} (\kvec)= (-v_{\alpha} ,u_{\alpha} )$
with
\begin{equation}
u^{2}_{\alpha}(\kvec) = \frac{1}{2} \bigg( 1 + \frac{\xi_{\alpha}}{E_{\alpha}} \bigg) \qquad 
v^{2}_{\alpha} (\kvec) = \frac{1}{2} \bigg( 1 - \frac{\xi_{\alpha}}{E_{\alpha}} \bigg)
\end{equation} 
where $\xi_{\alpha} = \epsilon_{\alpha}(\kk) - \mu $ and $E_{\alpha} = \sqrt{\xi_{\alpha}^2 + \Delta^2}$. The BdG eigenstates for the exciton condensate can be represented by the same expression with a different basis choice $(\kk,\sigma,e; -\kk,\sigma,h)$. This is due to the unified description allowed by the superconductor and exciton condensate.

For a $\kvec$-dependent and real order parameter $\Delta_{\kk}$, we can express the matrix element in the above expression as 
\begin{equation}
\langle \psi_{\alpha', +}| \partial_{\mu} H | \psi_{\alpha,-} \rangle =  - (v_{\alpha,\kk} u_{\alpha,\kk} + u_{\alpha',\kk} v_{\alpha,\kk}) \langle u_{\alpha'} | \partial_{\mu} H | u_{\alpha} \rangle.
\end{equation}
The matrix element above is calculated as usual 
\be 
\langle u_{\alpha'} | \partial_{\mu} H | u_{\alpha} \rangle = \partial_{\mu} \epsilon_{\alpha} \delta_{\alpha \alpha'} + (\epsilon_{\alpha'} (\kk) - \epsilon_{\alpha} (\kk)) \langle \partial_{\mu} u_{\alpha} | u_{\alpha'} \rangle. 
\ee
Taking the Fermi energy to lie within the $\alpha^{th}$ band, at zero temperature, a straightforward calculation~\cite{PhysRevB.95.024515} gives two contributions to the superfluid density for the $\alpha^{th}$-band $D_{\mu \nu} = D^{conv}_{\mu \nu} +D^{geo}_{\mu \nu}$, where,
\be
D^{conv}_{\mu \nu} =\frac{g}{L^2}\sum_{\kvec} \bigg( 1- \frac{\xi_{\alpha,\kk}}{E_{\alpha,\kk}} \partial_{\mu } \partial_{\nu} \epsilon_{\alpha,\kk} \bigg), \\
\ee
is the conventional contribution to superconductivity, while the geometric contribution $D^{geo}_{\mu \nu}$ becomes, 
\bea
D^{geo}_{\mu \nu} &=&\frac{g}{L^2}\sum_{\kvec} |\Delta_{\kk}|^2 \sum_{\alpha \neq \alpha'} \frac{(\xi_{\alpha}  - \xi_{\alpha'})}{\xi_{\alpha'}+\xi_{\alpha}} \bigg( \frac{1}{E_{\alpha'}} - \frac{1}{E_{\alpha}} \bigg) \nonumber \\
& &  \bigg[ \langle \partial_{\mu} u_{\alpha}  | u_{\alpha'} \rangle \langle u_{\alpha'} | \partial_{\mu} u_{\alpha} \rangle + h.c. \bigg], 
\eea
where $\alpha = \pm$ corresponds to the particle/hole bands of the chiral 2DEG. In the limit of a well-isolated band, the geometric superfluid can be expressed as~\cite{PhysRevB.95.024515}
\be
D^{geo}_{\mu \nu} = \frac{g}{L^2} \sum_{\kvec} 2 |\Delta_{\kk}|^2 \frac{1}{E_{\kk}} g_{\mu \nu} (\kk). 
\ee

\bibliographystyle{apsrev4-1}

\begin{thebibliography}{72}%
\makeatletter
\providecommand \@ifxundefined [1]{%
 \@ifx{#1\undefined}
}%
\providecommand \@ifnum [1]{%
 \ifnum #1\expandafter \@firstoftwo
 \else \expandafter \@secondoftwo
 \fi
}%
\providecommand \@ifx [1]{%
 \ifx #1\expandafter \@firstoftwo
 \else \expandafter \@secondoftwo
 \fi
}%
\providecommand \natexlab [1]{#1}%
\providecommand \enquote  [1]{``#1''}%
\providecommand \bibnamefont  [1]{#1}%
\providecommand \bibfnamefont [1]{#1}%
\providecommand \citenamefont [1]{#1}%
\providecommand \href@noop [0]{\@secondoftwo}%
\providecommand \href [0]{\begingroup \@sanitize@url \@href}%
\providecommand \@href[1]{\@@startlink{#1}\@@href}%
\providecommand \@@href[1]{\endgroup#1\@@endlink}%
\providecommand \@sanitize@url [0]{\catcode `\\12\catcode `\$12\catcode
  `\&12\catcode `\#12\catcode `\^12\catcode `\_12\catcode `\%12\relax}%
\providecommand \@@startlink[1]{}%
\providecommand \@@endlink[0]{}%
\providecommand \url  [0]{\begingroup\@sanitize@url \@url }%
\providecommand \@url [1]{\endgroup\@href {#1}{\urlprefix }}%
\providecommand \urlprefix  [0]{URL }%
\providecommand \Eprint [0]{\href }%
\providecommand \doibase [0]{http://dx.doi.org/}%
\providecommand \selectlanguage [0]{\@gobble}%
\providecommand \bibinfo  [0]{\@secondoftwo}%
\providecommand \bibfield  [0]{\@secondoftwo}%
\providecommand \translation [1]{[#1]}%
\providecommand \BibitemOpen [0]{}%
\providecommand \bibitemStop [0]{}%
\providecommand \bibitemNoStop [0]{.\EOS\space}%
\providecommand \EOS [0]{\spacefactor3000\relax}%
\providecommand \BibitemShut  [1]{\csname bibitem#1\endcsname}%
\let\auto@bib@innerbib\@empty
\bibitem [{\citenamefont {Cao}\ \emph {et~al.}(2018{\natexlab{a}})\citenamefont
  {Cao}, \citenamefont {Fatemi}, \citenamefont {Demir}, \citenamefont {Fang},
  \citenamefont {Tomarken}, \citenamefont {Luo}, \citenamefont
  {Sanchez-Yamagishi}, \citenamefont {Watanabe}, \citenamefont {Taniguchi},
  \citenamefont {Kaxiras} \emph {et~al.}}]{Cao2018}%
  \BibitemOpen
  \bibfield  {author} {\bibinfo {author} {\bibfnamefont {Y.}~\bibnamefont
  {Cao}}, \bibinfo {author} {\bibfnamefont {V.}~\bibnamefont {Fatemi}},
  \bibinfo {author} {\bibfnamefont {A.}~\bibnamefont {Demir}}, \bibinfo
  {author} {\bibfnamefont {S.}~\bibnamefont {Fang}}, \bibinfo {author}
  {\bibfnamefont {S.~L.}\ \bibnamefont {Tomarken}}, \bibinfo {author}
  {\bibfnamefont {J.~Y.}\ \bibnamefont {Luo}}, \bibinfo {author} {\bibfnamefont
  {J.~D.}\ \bibnamefont {Sanchez-Yamagishi}}, \bibinfo {author} {\bibfnamefont
  {K.}~\bibnamefont {Watanabe}}, \bibinfo {author} {\bibfnamefont
  {T.}~\bibnamefont {Taniguchi}}, \bibinfo {author} {\bibfnamefont
  {E.}~\bibnamefont {Kaxiras}},  \emph {et~al.},\ }\href@noop {} {\bibfield
  {journal} {\bibinfo  {journal} {Nature}\ }\textbf {\bibinfo {volume} {556}},\
  \bibinfo {pages} {80} (\bibinfo {year} {2018}{\natexlab{a}})}\BibitemShut
  {NoStop}%
\bibitem [{\citenamefont {Yankowitz}\ \emph {et~al.}(2019)\citenamefont
  {Yankowitz}, \citenamefont {Chen}, \citenamefont {Polshyn}, \citenamefont
  {Zhang}, \citenamefont {Watanabe}, \citenamefont {Taniguchi}, \citenamefont
  {Graf}, \citenamefont {Young},\ and\ \citenamefont {Dean}}]{Yankowitz2019}%
  \BibitemOpen
  \bibfield  {author} {\bibinfo {author} {\bibfnamefont {M.}~\bibnamefont
  {Yankowitz}}, \bibinfo {author} {\bibfnamefont {S.}~\bibnamefont {Chen}},
  \bibinfo {author} {\bibfnamefont {H.}~\bibnamefont {Polshyn}}, \bibinfo
  {author} {\bibfnamefont {Y.}~\bibnamefont {Zhang}}, \bibinfo {author}
  {\bibfnamefont {K.}~\bibnamefont {Watanabe}}, \bibinfo {author}
  {\bibfnamefont {T.}~\bibnamefont {Taniguchi}}, \bibinfo {author}
  {\bibfnamefont {D.}~\bibnamefont {Graf}}, \bibinfo {author} {\bibfnamefont
  {A.~F.}\ \bibnamefont {Young}}, \ and\ \bibinfo {author} {\bibfnamefont
  {C.~R.}\ \bibnamefont {Dean}},\ }\href@noop {} {\bibfield  {journal}
  {\bibinfo  {journal} {Science}\ }\textbf {\bibinfo {volume} {363}},\ \bibinfo
  {pages} {1059} (\bibinfo {year} {2019})}\BibitemShut {NoStop}%
\bibitem [{\citenamefont {Cao}\ \emph {et~al.}(2018{\natexlab{b}})\citenamefont
  {Cao}, \citenamefont {Fatemi}, \citenamefont {Fang}, \citenamefont
  {Watanabe}, \citenamefont {Taniguchi}, \citenamefont {Kaxiras},\ and\
  \citenamefont {Jarillo-Herrero}}]{Cao2018second}%
  \BibitemOpen
  \bibfield  {author} {\bibinfo {author} {\bibfnamefont {Y.}~\bibnamefont
  {Cao}}, \bibinfo {author} {\bibfnamefont {V.}~\bibnamefont {Fatemi}},
  \bibinfo {author} {\bibfnamefont {S.}~\bibnamefont {Fang}}, \bibinfo {author}
  {\bibfnamefont {K.}~\bibnamefont {Watanabe}}, \bibinfo {author}
  {\bibfnamefont {T.}~\bibnamefont {Taniguchi}}, \bibinfo {author}
  {\bibfnamefont {E.}~\bibnamefont {Kaxiras}}, \ and\ \bibinfo {author}
  {\bibfnamefont {P.}~\bibnamefont {Jarillo-Herrero}},\ }\href@noop {}
  {\bibfield  {journal} {\bibinfo  {journal} {Nature}\ }\textbf {\bibinfo
  {volume} {556}},\ \bibinfo {pages} {43} (\bibinfo {year}
  {2018}{\natexlab{b}})}\BibitemShut {NoStop}%
\bibitem [{\citenamefont {Lu}\ \emph {et~al.}(2019)\citenamefont {Lu},
  \citenamefont {Stepanov}, \citenamefont {Yang}, \citenamefont {Xie},
  \citenamefont {Aamir}, \citenamefont {Das}, \citenamefont {Urgell},
  \citenamefont {Watanabe}, \citenamefont {Taniguchi}, \citenamefont {Zhang}
  \emph {et~al.}}]{Lu2019}%
  \BibitemOpen
  \bibfield  {author} {\bibinfo {author} {\bibfnamefont {X.}~\bibnamefont
  {Lu}}, \bibinfo {author} {\bibfnamefont {P.}~\bibnamefont {Stepanov}},
  \bibinfo {author} {\bibfnamefont {W.}~\bibnamefont {Yang}}, \bibinfo {author}
  {\bibfnamefont {M.}~\bibnamefont {Xie}}, \bibinfo {author} {\bibfnamefont
  {M.~A.}\ \bibnamefont {Aamir}}, \bibinfo {author} {\bibfnamefont
  {I.}~\bibnamefont {Das}}, \bibinfo {author} {\bibfnamefont {C.}~\bibnamefont
  {Urgell}}, \bibinfo {author} {\bibfnamefont {K.}~\bibnamefont {Watanabe}},
  \bibinfo {author} {\bibfnamefont {T.}~\bibnamefont {Taniguchi}}, \bibinfo
  {author} {\bibfnamefont {G.}~\bibnamefont {Zhang}},  \emph {et~al.},\
  }\href@noop {} {\bibfield  {journal} {\bibinfo  {journal} {Nature}\ }\textbf
  {\bibinfo {volume} {574}},\ \bibinfo {pages} {653} (\bibinfo {year}
  {2019})}\BibitemShut {NoStop}%
\bibitem [{\citenamefont {Chen}\ \emph {et~al.}(2019)\citenamefont {Chen},
  \citenamefont {Sharpe}, \citenamefont {Gallagher}, \citenamefont {Rosen},
  \citenamefont {Fox}, \citenamefont {Jiang}, \citenamefont {Lyu},
  \citenamefont {Li}, \citenamefont {Watanabe}, \citenamefont {Taniguchi} \emph
  {et~al.}}]{Chen2019}%
  \BibitemOpen
  \bibfield  {author} {\bibinfo {author} {\bibfnamefont {G.}~\bibnamefont
  {Chen}}, \bibinfo {author} {\bibfnamefont {A.~L.}\ \bibnamefont {Sharpe}},
  \bibinfo {author} {\bibfnamefont {P.}~\bibnamefont {Gallagher}}, \bibinfo
  {author} {\bibfnamefont {I.~T.}\ \bibnamefont {Rosen}}, \bibinfo {author}
  {\bibfnamefont {E.~J.}\ \bibnamefont {Fox}}, \bibinfo {author} {\bibfnamefont
  {L.}~\bibnamefont {Jiang}}, \bibinfo {author} {\bibfnamefont
  {B.}~\bibnamefont {Lyu}}, \bibinfo {author} {\bibfnamefont {H.}~\bibnamefont
  {Li}}, \bibinfo {author} {\bibfnamefont {K.}~\bibnamefont {Watanabe}},
  \bibinfo {author} {\bibfnamefont {T.}~\bibnamefont {Taniguchi}},  \emph
  {et~al.},\ }\href@noop {} {\bibfield  {journal} {\bibinfo  {journal}
  {Nature}\ }\textbf {\bibinfo {volume} {572}},\ \bibinfo {pages} {215}
  (\bibinfo {year} {2019})}\BibitemShut {NoStop}%
\bibitem [{\citenamefont {Park}\ \emph {et~al.}(2021)\citenamefont {Park},
  \citenamefont {Cao}, \citenamefont {Watanabe}, \citenamefont {Taniguchi},\
  and\ \citenamefont {Jarillo-Herrero}}]{Park2021}%
  \BibitemOpen
  \bibfield  {author} {\bibinfo {author} {\bibfnamefont {J.~M.}\ \bibnamefont
  {Park}}, \bibinfo {author} {\bibfnamefont {Y.}~\bibnamefont {Cao}}, \bibinfo
  {author} {\bibfnamefont {K.}~\bibnamefont {Watanabe}}, \bibinfo {author}
  {\bibfnamefont {T.}~\bibnamefont {Taniguchi}}, \ and\ \bibinfo {author}
  {\bibfnamefont {P.}~\bibnamefont {Jarillo-Herrero}},\ }\href@noop {}
  {\bibfield  {journal} {\bibinfo  {journal} {Nature}\ }\textbf {\bibinfo
  {volume} {590}},\ \bibinfo {pages} {249} (\bibinfo {year}
  {2021})}\BibitemShut {NoStop}%
\bibitem [{\citenamefont {Cao}\ \emph {et~al.}(2021)\citenamefont {Cao},
  \citenamefont {Park}, \citenamefont {Watanabe}, \citenamefont {Taniguchi},\
  and\ \citenamefont {Jarillo-Herrero}}]{Cao2021}%
  \BibitemOpen
  \bibfield  {author} {\bibinfo {author} {\bibfnamefont {Y.}~\bibnamefont
  {Cao}}, \bibinfo {author} {\bibfnamefont {J.~M.}\ \bibnamefont {Park}},
  \bibinfo {author} {\bibfnamefont {K.}~\bibnamefont {Watanabe}}, \bibinfo
  {author} {\bibfnamefont {T.}~\bibnamefont {Taniguchi}}, \ and\ \bibinfo
  {author} {\bibfnamefont {P.}~\bibnamefont {Jarillo-Herrero}},\ }\href
  {\doibase 10.1038/s41586-021-03685-y} {\bibfield  {journal} {\bibinfo
  {journal} {Nature}\ }\textbf {\bibinfo {volume} {595}},\ \bibinfo {pages}
  {526} (\bibinfo {year} {2021})}\BibitemShut {NoStop}%
\bibitem [{\citenamefont {Sharpe}\ \emph {et~al.}(2019)\citenamefont {Sharpe},
  \citenamefont {Fox}, \citenamefont {Barnard}, \citenamefont {Finney},
  \citenamefont {Watanabe}, \citenamefont {Taniguchi}, \citenamefont
  {Kastner},\ and\ \citenamefont {Goldhaber-Gordon}}]{Sharpe605}%
  \BibitemOpen
  \bibfield  {author} {\bibinfo {author} {\bibfnamefont {A.~L.}\ \bibnamefont
  {Sharpe}}, \bibinfo {author} {\bibfnamefont {E.~J.}\ \bibnamefont {Fox}},
  \bibinfo {author} {\bibfnamefont {A.~W.}\ \bibnamefont {Barnard}}, \bibinfo
  {author} {\bibfnamefont {J.}~\bibnamefont {Finney}}, \bibinfo {author}
  {\bibfnamefont {K.}~\bibnamefont {Watanabe}}, \bibinfo {author}
  {\bibfnamefont {T.}~\bibnamefont {Taniguchi}}, \bibinfo {author}
  {\bibfnamefont {M.~A.}\ \bibnamefont {Kastner}}, \ and\ \bibinfo {author}
  {\bibfnamefont {D.}~\bibnamefont {Goldhaber-Gordon}},\ }\href {\doibase
  10.1126/science.aaw3780} {\bibfield  {journal} {\bibinfo  {journal}
  {Science}\ }\textbf {\bibinfo {volume} {365}},\ \bibinfo {pages} {605}
  (\bibinfo {year} {2019})}\BibitemShut {NoStop}%
\bibitem [{\citenamefont {Burg}\ \emph {et~al.}(2019)\citenamefont {Burg},
  \citenamefont {Zhu}, \citenamefont {Taniguchi}, \citenamefont {Watanabe},
  \citenamefont {MacDonald},\ and\ \citenamefont
  {Tutuc}}]{PhysRevLett.123.197702}%
  \BibitemOpen
  \bibfield  {author} {\bibinfo {author} {\bibfnamefont {G.~W.}\ \bibnamefont
  {Burg}}, \bibinfo {author} {\bibfnamefont {J.}~\bibnamefont {Zhu}}, \bibinfo
  {author} {\bibfnamefont {T.}~\bibnamefont {Taniguchi}}, \bibinfo {author}
  {\bibfnamefont {K.}~\bibnamefont {Watanabe}}, \bibinfo {author}
  {\bibfnamefont {A.~H.}\ \bibnamefont {MacDonald}}, \ and\ \bibinfo {author}
  {\bibfnamefont {E.}~\bibnamefont {Tutuc}},\ }\href {\doibase
  10.1103/PhysRevLett.123.197702} {\bibfield  {journal} {\bibinfo  {journal}
  {Phys. Rev. Lett.}\ }\textbf {\bibinfo {volume} {123}},\ \bibinfo {pages}
  {197702} (\bibinfo {year} {2019})}\BibitemShut {NoStop}%
\bibitem [{\citenamefont {Cao}\ \emph {et~al.}(2020)\citenamefont {Cao},
  \citenamefont {Rodan-Legrain}, \citenamefont {Rubies-Bigorda}, \citenamefont
  {Park}, \citenamefont {Watanabe}, \citenamefont {Taniguchi},\ and\
  \citenamefont {Jarillo-Herrero}}]{Cao2020}%
  \BibitemOpen
  \bibfield  {author} {\bibinfo {author} {\bibfnamefont {Y.}~\bibnamefont
  {Cao}}, \bibinfo {author} {\bibfnamefont {D.}~\bibnamefont {Rodan-Legrain}},
  \bibinfo {author} {\bibfnamefont {O.}~\bibnamefont {Rubies-Bigorda}},
  \bibinfo {author} {\bibfnamefont {J.~M.}\ \bibnamefont {Park}}, \bibinfo
  {author} {\bibfnamefont {K.}~\bibnamefont {Watanabe}}, \bibinfo {author}
  {\bibfnamefont {T.}~\bibnamefont {Taniguchi}}, \ and\ \bibinfo {author}
  {\bibfnamefont {P.}~\bibnamefont {Jarillo-Herrero}},\ }\href {\doibase
  10.1038/s41586-020-2260-6} {\bibfield  {journal} {\bibinfo  {journal}
  {Nature}\ }\textbf {\bibinfo {volume} {583}},\ \bibinfo {pages} {215}
  (\bibinfo {year} {2020})}\BibitemShut {NoStop}%
\bibitem [{\citenamefont {Polshyn}\ \emph {et~al.}(2020)\citenamefont
  {Polshyn}, \citenamefont {Zhu}, \citenamefont {Kumar}, \citenamefont {Zhang},
  \citenamefont {Yang}, \citenamefont {Tschirhart}, \citenamefont {Serlin},
  \citenamefont {Watanabe}, \citenamefont {Taniguchi}, \citenamefont
  {MacDonald},\ and\ \citenamefont {Young}}]{Polshyn_2020}%
  \BibitemOpen
  \bibfield  {author} {\bibinfo {author} {\bibfnamefont {H.}~\bibnamefont
  {Polshyn}}, \bibinfo {author} {\bibfnamefont {J.}~\bibnamefont {Zhu}},
  \bibinfo {author} {\bibfnamefont {M.~A.}\ \bibnamefont {Kumar}}, \bibinfo
  {author} {\bibfnamefont {Y.}~\bibnamefont {Zhang}}, \bibinfo {author}
  {\bibfnamefont {F.}~\bibnamefont {Yang}}, \bibinfo {author} {\bibfnamefont
  {C.~L.}\ \bibnamefont {Tschirhart}}, \bibinfo {author} {\bibfnamefont
  {M.}~\bibnamefont {Serlin}}, \bibinfo {author} {\bibfnamefont
  {K.}~\bibnamefont {Watanabe}}, \bibinfo {author} {\bibfnamefont
  {T.}~\bibnamefont {Taniguchi}}, \bibinfo {author} {\bibfnamefont {A.~H.}\
  \bibnamefont {MacDonald}}, \ and\ \bibinfo {author} {\bibfnamefont {A.~F.}\
  \bibnamefont {Young}},\ }\href {\doibase 10.1038/s41586-020-2963-8}
  {\bibfield  {journal} {\bibinfo  {journal} {Nature}\ }\textbf {\bibinfo
  {volume} {588}},\ \bibinfo {pages} {66} (\bibinfo {year} {2020})}\BibitemShut
  {NoStop}%
\bibitem [{\citenamefont {Chen}\ \emph {et~al.}(2021)\citenamefont {Chen},
  \citenamefont {He}, \citenamefont {Zhang}, \citenamefont {Hsieh},
  \citenamefont {Fei}, \citenamefont {Watanabe}, \citenamefont {Taniguchi},
  \citenamefont {Cobden}, \citenamefont {Xu}, \citenamefont {Dean},\ and\
  \citenamefont {Yankowitz}}]{Chen2021}%
  \BibitemOpen
  \bibfield  {author} {\bibinfo {author} {\bibfnamefont {S.}~\bibnamefont
  {Chen}}, \bibinfo {author} {\bibfnamefont {M.}~\bibnamefont {He}}, \bibinfo
  {author} {\bibfnamefont {Y.-H.}\ \bibnamefont {Zhang}}, \bibinfo {author}
  {\bibfnamefont {V.}~\bibnamefont {Hsieh}}, \bibinfo {author} {\bibfnamefont
  {Z.}~\bibnamefont {Fei}}, \bibinfo {author} {\bibfnamefont {K.}~\bibnamefont
  {Watanabe}}, \bibinfo {author} {\bibfnamefont {T.}~\bibnamefont {Taniguchi}},
  \bibinfo {author} {\bibfnamefont {D.~H.}\ \bibnamefont {Cobden}}, \bibinfo
  {author} {\bibfnamefont {X.}~\bibnamefont {Xu}}, \bibinfo {author}
  {\bibfnamefont {C.~R.}\ \bibnamefont {Dean}}, \ and\ \bibinfo {author}
  {\bibfnamefont {M.}~\bibnamefont {Yankowitz}},\ }\href {\doibase
  10.1038/s41567-020-01062-6} {\bibfield  {journal} {\bibinfo  {journal}
  {Nature Physics}\ }\textbf {\bibinfo {volume} {17}},\ \bibinfo {pages} {374}
  (\bibinfo {year} {2021})}\BibitemShut {NoStop}%
\bibitem [{\citenamefont {Xie}\ \emph {et~al.}(2021)\citenamefont {Xie},
  \citenamefont {Pierce}, \citenamefont {Park}, \citenamefont {Parker},
  \citenamefont {Khalaf}, \citenamefont {Ledwith}, \citenamefont {Cao},
  \citenamefont {Lee}, \citenamefont {Chen}, \citenamefont {Forrester},
  \citenamefont {Watanabe}, \citenamefont {Taniguchi}, \citenamefont
  {Vishwanath}, \citenamefont {Jarillo-Herrero},\ and\ \citenamefont
  {Yacoby}}]{Xie2021}%
  \BibitemOpen
  \bibfield  {author} {\bibinfo {author} {\bibfnamefont {Y.}~\bibnamefont
  {Xie}}, \bibinfo {author} {\bibfnamefont {A.~T.}\ \bibnamefont {Pierce}},
  \bibinfo {author} {\bibfnamefont {J.~M.}\ \bibnamefont {Park}}, \bibinfo
  {author} {\bibfnamefont {D.~E.}\ \bibnamefont {Parker}}, \bibinfo {author}
  {\bibfnamefont {E.}~\bibnamefont {Khalaf}}, \bibinfo {author} {\bibfnamefont
  {P.}~\bibnamefont {Ledwith}}, \bibinfo {author} {\bibfnamefont
  {Y.}~\bibnamefont {Cao}}, \bibinfo {author} {\bibfnamefont {S.~H.}\
  \bibnamefont {Lee}}, \bibinfo {author} {\bibfnamefont {S.}~\bibnamefont
  {Chen}}, \bibinfo {author} {\bibfnamefont {P.~R.}\ \bibnamefont {Forrester}},
  \bibinfo {author} {\bibfnamefont {K.}~\bibnamefont {Watanabe}}, \bibinfo
  {author} {\bibfnamefont {T.}~\bibnamefont {Taniguchi}}, \bibinfo {author}
  {\bibfnamefont {A.}~\bibnamefont {Vishwanath}}, \bibinfo {author}
  {\bibfnamefont {P.}~\bibnamefont {Jarillo-Herrero}}, \ and\ \bibinfo {author}
  {\bibfnamefont {A.}~\bibnamefont {Yacoby}},\ }\href {\doibase
  10.1038/s41586-021-04002-3} {\bibfield  {journal} {\bibinfo  {journal}
  {Nature}\ }\textbf {\bibinfo {volume} {600}},\ \bibinfo {pages} {439}
  (\bibinfo {year} {2021})}\BibitemShut {NoStop}%
\bibitem [{\citenamefont {Bistritzer}\ and\ \citenamefont
  {MacDonald}(2011)}]{bistritzer2011moire}%
  \BibitemOpen
  \bibfield  {author} {\bibinfo {author} {\bibfnamefont {R.}~\bibnamefont
  {Bistritzer}}\ and\ \bibinfo {author} {\bibfnamefont {A.~H.}\ \bibnamefont
  {MacDonald}},\ }\href@noop {} {\bibfield  {journal} {\bibinfo  {journal}
  {Proceedings of the National Academy of Sciences}\ }\textbf {\bibinfo
  {volume} {108}},\ \bibinfo {pages} {12233} (\bibinfo {year}
  {2011})}\BibitemShut {NoStop}%
\bibitem [{\citenamefont {Peotta}\ and\ \citenamefont
  {T{\"o}rm{\"a}}(2015)}]{Peotta2015}%
  \BibitemOpen
  \bibfield  {author} {\bibinfo {author} {\bibfnamefont {S.}~\bibnamefont
  {Peotta}}\ and\ \bibinfo {author} {\bibfnamefont {P.}~\bibnamefont
  {T{\"o}rm{\"a}}},\ }\href {\doibase 10.1038/ncomms9944} {\bibfield  {journal}
  {\bibinfo  {journal} {Nature Communications}\ }\textbf {\bibinfo {volume}
  {6}},\ \bibinfo {pages} {8944} (\bibinfo {year} {2015})}\BibitemShut
  {NoStop}%
\bibitem [{\citenamefont {Liang}\ \emph {et~al.}(2017)\citenamefont {Liang},
  \citenamefont {Vanhala}, \citenamefont {Peotta}, \citenamefont {Siro},
  \citenamefont {Harju},\ and\ \citenamefont {T\"orm\"a}}]{PhysRevB.95.024515}%
  \BibitemOpen
  \bibfield  {author} {\bibinfo {author} {\bibfnamefont {L.}~\bibnamefont
  {Liang}}, \bibinfo {author} {\bibfnamefont {T.~I.}\ \bibnamefont {Vanhala}},
  \bibinfo {author} {\bibfnamefont {S.}~\bibnamefont {Peotta}}, \bibinfo
  {author} {\bibfnamefont {T.}~\bibnamefont {Siro}}, \bibinfo {author}
  {\bibfnamefont {A.}~\bibnamefont {Harju}}, \ and\ \bibinfo {author}
  {\bibfnamefont {P.}~\bibnamefont {T\"orm\"a}},\ }\href {\doibase
  10.1103/PhysRevB.95.024515} {\bibfield  {journal} {\bibinfo  {journal} {Phys.
  Rev. B}\ }\textbf {\bibinfo {volume} {95}},\ \bibinfo {pages} {024515}
  (\bibinfo {year} {2017})}\BibitemShut {NoStop}%
\bibitem [{\citenamefont {Hu}\ \emph {et~al.}(2019)\citenamefont {Hu},
  \citenamefont {Hyart}, \citenamefont {Pikulin},\ and\ \citenamefont
  {Rossi}}]{PhysRevLett.123.237002}%
  \BibitemOpen
  \bibfield  {author} {\bibinfo {author} {\bibfnamefont {X.}~\bibnamefont
  {Hu}}, \bibinfo {author} {\bibfnamefont {T.}~\bibnamefont {Hyart}}, \bibinfo
  {author} {\bibfnamefont {D.~I.}\ \bibnamefont {Pikulin}}, \ and\ \bibinfo
  {author} {\bibfnamefont {E.}~\bibnamefont {Rossi}},\ }\href {\doibase
  10.1103/PhysRevLett.123.237002} {\bibfield  {journal} {\bibinfo  {journal}
  {Phys. Rev. Lett.}\ }\textbf {\bibinfo {volume} {123}},\ \bibinfo {pages}
  {237002} (\bibinfo {year} {2019})}\BibitemShut {NoStop}%
\bibitem [{\citenamefont {Xie}\ \emph {et~al.}(2020)\citenamefont {Xie},
  \citenamefont {Song}, \citenamefont {Lian},\ and\ \citenamefont
  {Bernevig}}]{PhysRevLett.124.167002}%
  \BibitemOpen
  \bibfield  {author} {\bibinfo {author} {\bibfnamefont {F.}~\bibnamefont
  {Xie}}, \bibinfo {author} {\bibfnamefont {Z.}~\bibnamefont {Song}}, \bibinfo
  {author} {\bibfnamefont {B.}~\bibnamefont {Lian}}, \ and\ \bibinfo {author}
  {\bibfnamefont {B.~A.}\ \bibnamefont {Bernevig}},\ }\href {\doibase
  10.1103/PhysRevLett.124.167002} {\bibfield  {journal} {\bibinfo  {journal}
  {Phys. Rev. Lett.}\ }\textbf {\bibinfo {volume} {124}},\ \bibinfo {pages}
  {167002} (\bibinfo {year} {2020})}\BibitemShut {NoStop}%
\bibitem [{\citenamefont {Julku}\ \emph {et~al.}(2020)\citenamefont {Julku},
  \citenamefont {Peltonen}, \citenamefont {Liang}, \citenamefont {Heikkil\"a},\
  and\ \citenamefont {T\"orm\"a}}]{PhysRevB.101.060505}%
  \BibitemOpen
  \bibfield  {author} {\bibinfo {author} {\bibfnamefont {A.}~\bibnamefont
  {Julku}}, \bibinfo {author} {\bibfnamefont {T.~J.}\ \bibnamefont {Peltonen}},
  \bibinfo {author} {\bibfnamefont {L.}~\bibnamefont {Liang}}, \bibinfo
  {author} {\bibfnamefont {T.~T.}\ \bibnamefont {Heikkil\"a}}, \ and\ \bibinfo
  {author} {\bibfnamefont {P.}~\bibnamefont {T\"orm\"a}},\ }\href {\doibase
  10.1103/PhysRevB.101.060505} {\bibfield  {journal} {\bibinfo  {journal}
  {Phys. Rev. B}\ }\textbf {\bibinfo {volume} {101}},\ \bibinfo {pages}
  {060505} (\bibinfo {year} {2020})}\BibitemShut {NoStop}%
\bibitem [{\citenamefont {Julku}\ \emph {et~al.}(2021)\citenamefont {Julku},
  \citenamefont {Bruun},\ and\ \citenamefont
  {T{\"o}rm{\"a}}}]{PhysRevLett.127.170404}%
  \BibitemOpen
  \bibfield  {author} {\bibinfo {author} {\bibfnamefont {A.}~\bibnamefont
  {Julku}}, \bibinfo {author} {\bibfnamefont {G.~M.}\ \bibnamefont {Bruun}}, \
  and\ \bibinfo {author} {\bibfnamefont {P.}~\bibnamefont {T{\"o}rm{\"a}}},\
  }\href@noop {} {\bibfield  {journal} {\bibinfo  {journal} {Phys. Rev. Lett.}\
  }\textbf {\bibinfo {volume} {127}},\ \bibinfo {pages} {170404} (\bibinfo
  {year} {2021})}\BibitemShut {NoStop}%
\bibitem [{\citenamefont {T{\"o}rm{\"a}}\ \emph {et~al.}(2018)\citenamefont
  {T{\"o}rm{\"a}}, \citenamefont {Liang},\ and\ \citenamefont
  {Peotta}}]{PhysRevB.98.220511}%
  \BibitemOpen
  \bibfield  {author} {\bibinfo {author} {\bibfnamefont {P.}~\bibnamefont
  {T{\"o}rm{\"a}}}, \bibinfo {author} {\bibfnamefont {L.}~\bibnamefont
  {Liang}}, \ and\ \bibinfo {author} {\bibfnamefont {S.}~\bibnamefont
  {Peotta}},\ }\href@noop {} {\bibfield  {journal} {\bibinfo  {journal}
  {Physical Review B}\ }\textbf {\bibinfo {volume} {98}},\ \bibinfo {pages}
  {220511} (\bibinfo {year} {2018})}\BibitemShut {NoStop}%
\bibitem [{\citenamefont {Wang}\ \emph {et~al.}(2020)\citenamefont {Wang},
  \citenamefont {Chaudhary}, \citenamefont {Chen},\ and\ \citenamefont
  {Levin}}]{PhysRevB.102.184504}%
  \BibitemOpen
  \bibfield  {author} {\bibinfo {author} {\bibfnamefont {Z.}~\bibnamefont
  {Wang}}, \bibinfo {author} {\bibfnamefont {G.}~\bibnamefont {Chaudhary}},
  \bibinfo {author} {\bibfnamefont {Q.}~\bibnamefont {Chen}}, \ and\ \bibinfo
  {author} {\bibfnamefont {K.}~\bibnamefont {Levin}},\ }\href@noop {}
  {\bibfield  {journal} {\bibinfo  {journal} {Physical Review B}\ }\textbf
  {\bibinfo {volume} {102}},\ \bibinfo {pages} {184504} (\bibinfo {year}
  {2020})}\BibitemShut {NoStop}%
\bibitem [{\citenamefont {Julku}\ \emph {et~al.}(2016)\citenamefont {Julku},
  \citenamefont {Peotta}, \citenamefont {Vanhala}, \citenamefont {Kim},\ and\
  \citenamefont {T{\"o}rm{\"a}}}]{PhysRevLett.117.045303}%
  \BibitemOpen
  \bibfield  {author} {\bibinfo {author} {\bibfnamefont {A.}~\bibnamefont
  {Julku}}, \bibinfo {author} {\bibfnamefont {S.}~\bibnamefont {Peotta}},
  \bibinfo {author} {\bibfnamefont {T.~I.}\ \bibnamefont {Vanhala}}, \bibinfo
  {author} {\bibfnamefont {D.-H.}\ \bibnamefont {Kim}}, \ and\ \bibinfo
  {author} {\bibfnamefont {P.}~\bibnamefont {T{\"o}rm{\"a}}},\ }\href@noop {}
  {\bibfield  {journal} {\bibinfo  {journal} {Phys. Rev. Lett.}\ }\textbf
  {\bibinfo {volume} {117}},\ \bibinfo {pages} {045303} (\bibinfo {year}
  {2016})}\BibitemShut {NoStop}%
\bibitem [{\citenamefont {Rossi}(2021)}]{Rossi2021quantum}%
  \BibitemOpen
  \bibfield  {author} {\bibinfo {author} {\bibfnamefont {E.}~\bibnamefont
  {Rossi}},\ }\href {\doibase https://doi.org/10.1016/j.cossms.2021.100952}
  {\bibfield  {journal} {\bibinfo  {journal} {Current Opinion in Solid State
  and Materials Science}\ }\textbf {\bibinfo {volume} {25}},\ \bibinfo {pages}
  {100952} (\bibinfo {year} {2021})}\BibitemShut {NoStop}%
\bibitem [{\citenamefont {Parameswaran}\ \emph {et~al.}(2013)\citenamefont
  {Parameswaran}, \citenamefont {Roy},\ and\ \citenamefont
  {Sondhi}}]{Roysondhi2013816}%
  \BibitemOpen
  \bibfield  {author} {\bibinfo {author} {\bibfnamefont {S.~A.}\ \bibnamefont
  {Parameswaran}}, \bibinfo {author} {\bibfnamefont {R.}~\bibnamefont {Roy}}, \
  and\ \bibinfo {author} {\bibfnamefont {S.~L.}\ \bibnamefont {Sondhi}},\
  }\href {\doibase https://doi.org/10.1016/j.crhy.2013.04.003} {\bibfield
  {journal} {\bibinfo  {journal} {Comptes Rendus Physique}\ }\textbf {\bibinfo
  {volume} {14}},\ \bibinfo {pages} {816 } (\bibinfo {year} {2013})},\ \bibinfo
  {note} {topological insulators / Isolants topologiques}\BibitemShut {NoStop}%
\bibitem [{\citenamefont {Roy}(2014)}]{PhysRevB.90.165139}%
  \BibitemOpen
  \bibfield  {author} {\bibinfo {author} {\bibfnamefont {R.}~\bibnamefont
  {Roy}},\ }\href {\doibase 10.1103/PhysRevB.90.165139} {\bibfield  {journal}
  {\bibinfo  {journal} {Phys. Rev. B}\ }\textbf {\bibinfo {volume} {90}},\
  \bibinfo {pages} {165139} (\bibinfo {year} {2014})}\BibitemShut {NoStop}%
\bibitem [{\citenamefont {Haldane}(2011)}]{PhysRevLett.107.116801}%
  \BibitemOpen
  \bibfield  {author} {\bibinfo {author} {\bibfnamefont {F.~D.~M.}\
  \bibnamefont {Haldane}},\ }\href {\doibase 10.1103/PhysRevLett.107.116801}
  {\bibfield  {journal} {\bibinfo  {journal} {Phys. Rev. Lett.}\ }\textbf
  {\bibinfo {volume} {107}},\ \bibinfo {pages} {116801} (\bibinfo {year}
  {2011})}\BibitemShut {NoStop}%
\bibitem [{\citenamefont {Parameswaran}\ \emph {et~al.}(2012)\citenamefont
  {Parameswaran}, \citenamefont {Roy},\ and\ \citenamefont
  {Sondhi}}]{PhysRevB.85.241308}%
  \BibitemOpen
  \bibfield  {author} {\bibinfo {author} {\bibfnamefont {S.~A.}\ \bibnamefont
  {Parameswaran}}, \bibinfo {author} {\bibfnamefont {R.}~\bibnamefont {Roy}}, \
  and\ \bibinfo {author} {\bibfnamefont {S.~L.}\ \bibnamefont {Sondhi}},\
  }\href {\doibase 10.1103/PhysRevB.85.241308} {\bibfield  {journal} {\bibinfo
  {journal} {Phys. Rev. B}\ }\textbf {\bibinfo {volume} {85}},\ \bibinfo
  {pages} {241308} (\bibinfo {year} {2012})}\BibitemShut {NoStop}%
\bibitem [{\citenamefont {Regnault}\ and\ \citenamefont
  {Bernevig}(2011)}]{PhysRevX.1.021014}%
  \BibitemOpen
  \bibfield  {author} {\bibinfo {author} {\bibfnamefont {N.}~\bibnamefont
  {Regnault}}\ and\ \bibinfo {author} {\bibfnamefont {B.~A.}\ \bibnamefont
  {Bernevig}},\ }\href {\doibase 10.1103/PhysRevX.1.021014} {\bibfield
  {journal} {\bibinfo  {journal} {Phys. Rev. X}\ }\textbf {\bibinfo {volume}
  {1}},\ \bibinfo {pages} {021014} (\bibinfo {year} {2011})}\BibitemShut
  {NoStop}%
\bibitem [{\citenamefont {Jackson}\ \emph {et~al.}(2015)\citenamefont
  {Jackson}, \citenamefont {M{\"o}ller},\ and\ \citenamefont
  {Roy}}]{Jackson2015}%
  \BibitemOpen
  \bibfield  {author} {\bibinfo {author} {\bibfnamefont {T.~S.}\ \bibnamefont
  {Jackson}}, \bibinfo {author} {\bibfnamefont {G.}~\bibnamefont {M{\"o}ller}},
  \ and\ \bibinfo {author} {\bibfnamefont {R.}~\bibnamefont {Roy}},\ }\href
  {\doibase 10.1038/ncomms9629} {\bibfield  {journal} {\bibinfo  {journal}
  {Nature Communications}\ }\textbf {\bibinfo {volume} {6}},\ \bibinfo {pages}
  {8629} (\bibinfo {year} {2015})}\BibitemShut {NoStop}%
\bibitem [{\citenamefont {Andrews}\ and\ \citenamefont
  {Soluyanov}(2020)}]{PhysRevB.101.235312}%
  \BibitemOpen
  \bibfield  {author} {\bibinfo {author} {\bibfnamefont {B.}~\bibnamefont
  {Andrews}}\ and\ \bibinfo {author} {\bibfnamefont {A.}~\bibnamefont
  {Soluyanov}},\ }\href {\doibase 10.1103/PhysRevB.101.235312} {\bibfield
  {journal} {\bibinfo  {journal} {Phys. Rev. B}\ }\textbf {\bibinfo {volume}
  {101}},\ \bibinfo {pages} {235312} (\bibinfo {year} {2020})}\BibitemShut
  {NoStop}%
\bibitem [{\citenamefont {Repellin}\ and\ \citenamefont
  {Senthil}(2020)}]{PhysRevResearch.2.023238}%
  \BibitemOpen
  \bibfield  {author} {\bibinfo {author} {\bibfnamefont {C.}~\bibnamefont
  {Repellin}}\ and\ \bibinfo {author} {\bibfnamefont {T.}~\bibnamefont
  {Senthil}},\ }\href {\doibase 10.1103/PhysRevResearch.2.023238} {\bibfield
  {journal} {\bibinfo  {journal} {Phys. Rev. Res.}\ }\textbf {\bibinfo {volume}
  {2}},\ \bibinfo {pages} {023238} (\bibinfo {year} {2020})}\BibitemShut
  {NoStop}%
\bibitem [{\citenamefont {Ledwith}\ \emph {et~al.}(2020)\citenamefont
  {Ledwith}, \citenamefont {Tarnopolsky}, \citenamefont {Khalaf},\ and\
  \citenamefont {Vishwanath}}]{PhysRevResearch.2.023237}%
  \BibitemOpen
  \bibfield  {author} {\bibinfo {author} {\bibfnamefont {P.~J.}\ \bibnamefont
  {Ledwith}}, \bibinfo {author} {\bibfnamefont {G.}~\bibnamefont
  {Tarnopolsky}}, \bibinfo {author} {\bibfnamefont {E.}~\bibnamefont {Khalaf}},
  \ and\ \bibinfo {author} {\bibfnamefont {A.}~\bibnamefont {Vishwanath}},\
  }\href {\doibase 10.1103/PhysRevResearch.2.023237} {\bibfield  {journal}
  {\bibinfo  {journal} {Phys. Rev. Res.}\ }\textbf {\bibinfo {volume} {2}},\
  \bibinfo {pages} {023237} (\bibinfo {year} {2020})}\BibitemShut {NoStop}%
\bibitem [{\citenamefont {Wilhelm}\ \emph {et~al.}(2021)\citenamefont
  {Wilhelm}, \citenamefont {Lang},\ and\ \citenamefont
  {L\"auchli}}]{PhysRevB.103.125406}%
  \BibitemOpen
  \bibfield  {author} {\bibinfo {author} {\bibfnamefont {P.}~\bibnamefont
  {Wilhelm}}, \bibinfo {author} {\bibfnamefont {T.~C.}\ \bibnamefont {Lang}}, \
  and\ \bibinfo {author} {\bibfnamefont {A.~M.}\ \bibnamefont {L\"auchli}},\
  }\href {\doibase 10.1103/PhysRevB.103.125406} {\bibfield  {journal} {\bibinfo
   {journal} {Phys. Rev. B}\ }\textbf {\bibinfo {volume} {103}},\ \bibinfo
  {pages} {125406} (\bibinfo {year} {2021})}\BibitemShut {NoStop}%
\bibitem [{\citenamefont {Kwan}\ \emph {et~al.}(2021)\citenamefont {Kwan},
  \citenamefont {Hu}, \citenamefont {Simon},\ and\ \citenamefont
  {Parameswaran}}]{PhysRevLett.126.137601}%
  \BibitemOpen
  \bibfield  {author} {\bibinfo {author} {\bibfnamefont {Y.~H.}\ \bibnamefont
  {Kwan}}, \bibinfo {author} {\bibfnamefont {Y.}~\bibnamefont {Hu}}, \bibinfo
  {author} {\bibfnamefont {S.~H.}\ \bibnamefont {Simon}}, \ and\ \bibinfo
  {author} {\bibfnamefont {S.~A.}\ \bibnamefont {Parameswaran}},\ }\href
  {\doibase 10.1103/PhysRevLett.126.137601} {\bibfield  {journal} {\bibinfo
  {journal} {Phys. Rev. Lett.}\ }\textbf {\bibinfo {volume} {126}},\ \bibinfo
  {pages} {137601} (\bibinfo {year} {2021})}\BibitemShut {NoStop}%
\bibitem [{\citenamefont {Hu}\ \emph {et~al.}(2022)\citenamefont {Hu},
  \citenamefont {Hyart}, \citenamefont {Pikulin},\ and\ \citenamefont
  {Rossi}}]{PhysRevB.105.L140506}%
  \BibitemOpen
  \bibfield  {author} {\bibinfo {author} {\bibfnamefont {X.}~\bibnamefont
  {Hu}}, \bibinfo {author} {\bibfnamefont {T.}~\bibnamefont {Hyart}}, \bibinfo
  {author} {\bibfnamefont {D.~I.}\ \bibnamefont {Pikulin}}, \ and\ \bibinfo
  {author} {\bibfnamefont {E.}~\bibnamefont {Rossi}},\ }\href {\doibase
  10.1103/PhysRevB.105.L140506} {\bibfield  {journal} {\bibinfo  {journal}
  {Physical Review B}\ }\textbf {\bibinfo {volume} {105}},\ \bibinfo {pages}
  {L140506} (\bibinfo {year} {2022})}\BibitemShut {NoStop}%
\bibitem [{\citenamefont {Schrieffer}(1999)}]{schrieffer1999theory}%
  \BibitemOpen
  \bibfield  {author} {\bibinfo {author} {\bibfnamefont {J.}~\bibnamefont
  {Schrieffer}},\ }\href@noop {} {\emph {\bibinfo {title} {Theory Of
  Superconductivity}}},\ Advanced Books Classics\ (\bibinfo  {publisher}
  {Avalon Publishing},\ \bibinfo {year} {1999})\BibitemShut {NoStop}%
\bibitem [{\citenamefont {Wang}\ \emph {et~al.}(2019)\citenamefont {Wang},
  \citenamefont {Rhodes}, \citenamefont {Watanabe}, \citenamefont {Taniguchi},
  \citenamefont {Hone}, \citenamefont {Shan},\ and\ \citenamefont
  {Mak}}]{Wang2019}%
  \BibitemOpen
  \bibfield  {author} {\bibinfo {author} {\bibfnamefont {Z.}~\bibnamefont
  {Wang}}, \bibinfo {author} {\bibfnamefont {D.~A.}\ \bibnamefont {Rhodes}},
  \bibinfo {author} {\bibfnamefont {K.}~\bibnamefont {Watanabe}}, \bibinfo
  {author} {\bibfnamefont {T.}~\bibnamefont {Taniguchi}}, \bibinfo {author}
  {\bibfnamefont {J.~C.}\ \bibnamefont {Hone}}, \bibinfo {author}
  {\bibfnamefont {J.}~\bibnamefont {Shan}}, \ and\ \bibinfo {author}
  {\bibfnamefont {K.~F.}\ \bibnamefont {Mak}},\ }\href {\doibase
  10.1038/s41586-019-1591-7} {\bibfield  {journal} {\bibinfo  {journal}
  {Nature}\ }\textbf {\bibinfo {volume} {574}},\ \bibinfo {pages} {76}
  (\bibinfo {year} {2019})}\BibitemShut {NoStop}%
\bibitem [{\citenamefont {Herzog-Arbeitman}\ \emph {et~al.}(2022)\citenamefont
  {Herzog-Arbeitman}, \citenamefont {Peri}, \citenamefont {Schindler},
  \citenamefont {Huber},\ and\ \citenamefont
  {Bernevig}}]{PhysRevLett.128.087002}%
  \BibitemOpen
  \bibfield  {author} {\bibinfo {author} {\bibfnamefont {J.}~\bibnamefont
  {Herzog-Arbeitman}}, \bibinfo {author} {\bibfnamefont {V.}~\bibnamefont
  {Peri}}, \bibinfo {author} {\bibfnamefont {F.}~\bibnamefont {Schindler}},
  \bibinfo {author} {\bibfnamefont {S.~D.}\ \bibnamefont {Huber}}, \ and\
  \bibinfo {author} {\bibfnamefont {B.~A.}\ \bibnamefont {Bernevig}},\
  }\href@noop {} {\bibfield  {journal} {\bibinfo  {journal} {Phys. Rev. Lett.}\
  }\textbf {\bibinfo {volume} {128}},\ \bibinfo {pages} {087002} (\bibinfo
  {year} {2022})}\BibitemShut {NoStop}%
\bibitem [{\citenamefont {Tovmasyan}\ \emph {et~al.}(2016)\citenamefont
  {Tovmasyan}, \citenamefont {Peotta}, \citenamefont {T{\"o}rm{\"a}},\ and\
  \citenamefont {Huber}}]{PhysRevB.94.254149}%
  \BibitemOpen
  \bibfield  {author} {\bibinfo {author} {\bibfnamefont {M.}~\bibnamefont
  {Tovmasyan}}, \bibinfo {author} {\bibfnamefont {S.}~\bibnamefont {Peotta}},
  \bibinfo {author} {\bibfnamefont {P.}~\bibnamefont {T{\"o}rm{\"a}}}, \ and\
  \bibinfo {author} {\bibfnamefont {S.~D.}\ \bibnamefont {Huber}},\ }\href@noop
  {} {\bibfield  {journal} {\bibinfo  {journal} {Physical Review B}\ }\textbf
  {\bibinfo {volume} {94}},\ \bibinfo {pages} {245149} (\bibinfo {year}
  {2016})}\BibitemShut {NoStop}%
\bibitem [{\citenamefont {Hofmann}\ \emph {et~al.}(2020)\citenamefont
  {Hofmann}, \citenamefont {Berg},\ and\ \citenamefont
  {Chowdhury}}]{PhysRevB.102.201112}%
  \BibitemOpen
  \bibfield  {author} {\bibinfo {author} {\bibfnamefont {J.~S.}\ \bibnamefont
  {Hofmann}}, \bibinfo {author} {\bibfnamefont {E.}~\bibnamefont {Berg}}, \
  and\ \bibinfo {author} {\bibfnamefont {D.}~\bibnamefont {Chowdhury}},\
  }\href@noop {} {\bibfield  {journal} {\bibinfo  {journal} {Physical Review
  B}\ }\textbf {\bibinfo {volume} {102}},\ \bibinfo {pages} {201112} (\bibinfo
  {year} {2020})}\BibitemShut {NoStop}%
\bibitem [{\citenamefont {Tian}\ \emph {et~al.}(2023)\citenamefont {Tian},
  \citenamefont {Gao}, \citenamefont {Zhang}, \citenamefont {Che},
  \citenamefont {Xu}, \citenamefont {Cheung}, \citenamefont {Watanabe},
  \citenamefont {Taniguchi}, \citenamefont {Randeria}, \citenamefont {Zhang},
  \citenamefont {Lau},\ and\ \citenamefont {Bockrath}}]{Tian2023}%
  \BibitemOpen
  \bibfield  {author} {\bibinfo {author} {\bibfnamefont {H.}~\bibnamefont
  {Tian}}, \bibinfo {author} {\bibfnamefont {X.}~\bibnamefont {Gao}}, \bibinfo
  {author} {\bibfnamefont {Y.}~\bibnamefont {Zhang}}, \bibinfo {author}
  {\bibfnamefont {S.}~\bibnamefont {Che}}, \bibinfo {author} {\bibfnamefont
  {T.}~\bibnamefont {Xu}}, \bibinfo {author} {\bibfnamefont {P.}~\bibnamefont
  {Cheung}}, \bibinfo {author} {\bibfnamefont {K.}~\bibnamefont {Watanabe}},
  \bibinfo {author} {\bibfnamefont {T.}~\bibnamefont {Taniguchi}}, \bibinfo
  {author} {\bibfnamefont {M.}~\bibnamefont {Randeria}}, \bibinfo {author}
  {\bibfnamefont {F.}~\bibnamefont {Zhang}}, \bibinfo {author} {\bibfnamefont
  {C.~N.}\ \bibnamefont {Lau}}, \ and\ \bibinfo {author} {\bibfnamefont
  {M.~W.}\ \bibnamefont {Bockrath}},\ }\href {\doibase
  10.1038/s41586-022-05576-2} {\bibfield  {journal} {\bibinfo  {journal}
  {Nature}\ }\textbf {\bibinfo {volume} {614}},\ \bibinfo {pages} {440}
  (\bibinfo {year} {2023})}\BibitemShut {NoStop}%
\bibitem [{\citenamefont {McCann}\ and\ \citenamefont
  {Fal'ko}(2006)}]{PhysRevLett.96.086805}%
  \BibitemOpen
  \bibfield  {author} {\bibinfo {author} {\bibfnamefont {E.}~\bibnamefont
  {McCann}}\ and\ \bibinfo {author} {\bibfnamefont {V.~I.}\ \bibnamefont
  {Fal'ko}},\ }\href {\doibase 10.1103/PhysRevLett.96.086805} {\bibfield
  {journal} {\bibinfo  {journal} {Phys. Rev. Lett.}\ }\textbf {\bibinfo
  {volume} {96}},\ \bibinfo {pages} {086805} (\bibinfo {year}
  {2006})}\BibitemShut {NoStop}%
\bibitem [{\citenamefont {Min}\ and\ \citenamefont
  {MacDonald}(2008)}]{PhysRevB.77.155416}%
  \BibitemOpen
  \bibfield  {author} {\bibinfo {author} {\bibfnamefont {H.}~\bibnamefont
  {Min}}\ and\ \bibinfo {author} {\bibfnamefont {A.~H.}\ \bibnamefont
  {MacDonald}},\ }\href {\doibase 10.1103/PhysRevB.77.155416} {\bibfield
  {journal} {\bibinfo  {journal} {Phys. Rev. B}\ }\textbf {\bibinfo {volume}
  {77}},\ \bibinfo {pages} {155416} (\bibinfo {year} {2008})}\BibitemShut
  {NoStop}%
\bibitem [{\citenamefont {Guinea}\ \emph {et~al.}(2006)\citenamefont {Guinea},
  \citenamefont {Castro~Neto},\ and\ \citenamefont
  {Peres}}]{PhysRevB.73.245426}%
  \BibitemOpen
  \bibfield  {author} {\bibinfo {author} {\bibfnamefont {F.}~\bibnamefont
  {Guinea}}, \bibinfo {author} {\bibfnamefont {A.~H.}\ \bibnamefont
  {Castro~Neto}}, \ and\ \bibinfo {author} {\bibfnamefont {N.~M.~R.}\
  \bibnamefont {Peres}},\ }\href {\doibase 10.1103/PhysRevB.73.245426}
  {\bibfield  {journal} {\bibinfo  {journal} {Phys. Rev. B}\ }\textbf {\bibinfo
  {volume} {73}},\ \bibinfo {pages} {245426} (\bibinfo {year}
  {2006})}\BibitemShut {NoStop}%
\bibitem [{\citenamefont {Koshino}\ and\ \citenamefont
  {McCann}(2010)}]{PhysRevB.81.115315}%
  \BibitemOpen
  \bibfield  {author} {\bibinfo {author} {\bibfnamefont {M.}~\bibnamefont
  {Koshino}}\ and\ \bibinfo {author} {\bibfnamefont {E.}~\bibnamefont
  {McCann}},\ }\href {\doibase 10.1103/PhysRevB.81.115315} {\bibfield
  {journal} {\bibinfo  {journal} {Phys. Rev. B}\ }\textbf {\bibinfo {volume}
  {81}},\ \bibinfo {pages} {115315} (\bibinfo {year} {2010})}\BibitemShut
  {NoStop}%
\bibitem [{\citenamefont {Serbyn}\ and\ \citenamefont
  {Abanin}(2013)}]{PhysRevB.87.115422}%
  \BibitemOpen
  \bibfield  {author} {\bibinfo {author} {\bibfnamefont {M.}~\bibnamefont
  {Serbyn}}\ and\ \bibinfo {author} {\bibfnamefont {D.~A.}\ \bibnamefont
  {Abanin}},\ }\href {\doibase 10.1103/PhysRevB.87.115422} {\bibfield
  {journal} {\bibinfo  {journal} {Phys. Rev. B}\ }\textbf {\bibinfo {volume}
  {87}},\ \bibinfo {pages} {115422} (\bibinfo {year} {2013})}\BibitemShut
  {NoStop}%
\bibitem [{\citenamefont {Koshino}\ and\ \citenamefont
  {McCann}(2011)}]{PhysRevB.83.165443}%
  \BibitemOpen
  \bibfield  {author} {\bibinfo {author} {\bibfnamefont {M.}~\bibnamefont
  {Koshino}}\ and\ \bibinfo {author} {\bibfnamefont {E.}~\bibnamefont
  {McCann}},\ }\href {\doibase 10.1103/PhysRevB.83.165443} {\bibfield
  {journal} {\bibinfo  {journal} {Phys. Rev. B}\ }\textbf {\bibinfo {volume}
  {83}},\ \bibinfo {pages} {165443} (\bibinfo {year} {2011})}\BibitemShut
  {NoStop}%
\bibitem [{\citenamefont {Zhou}\ \emph {et~al.}(2022)\citenamefont {Zhou},
  \citenamefont {Holleis}, \citenamefont {Saito}, \citenamefont {Cohen},
  \citenamefont {Huynh}, \citenamefont {Patterson}, \citenamefont {Yang},
  \citenamefont {Taniguchi}, \citenamefont {Watanabe},\ and\ \citenamefont
  {Young}}]{BLGsuperconductivity}%
  \BibitemOpen
  \bibfield  {author} {\bibinfo {author} {\bibfnamefont {H.}~\bibnamefont
  {Zhou}}, \bibinfo {author} {\bibfnamefont {L.}~\bibnamefont {Holleis}},
  \bibinfo {author} {\bibfnamefont {Y.}~\bibnamefont {Saito}}, \bibinfo
  {author} {\bibfnamefont {L.}~\bibnamefont {Cohen}}, \bibinfo {author}
  {\bibfnamefont {W.}~\bibnamefont {Huynh}}, \bibinfo {author} {\bibfnamefont
  {C.~L.}\ \bibnamefont {Patterson}}, \bibinfo {author} {\bibfnamefont
  {F.}~\bibnamefont {Yang}}, \bibinfo {author} {\bibfnamefont {T.}~\bibnamefont
  {Taniguchi}}, \bibinfo {author} {\bibfnamefont {K.}~\bibnamefont {Watanabe}},
  \ and\ \bibinfo {author} {\bibfnamefont {A.~F.}\ \bibnamefont {Young}},\
  }\href {\doibase 10.1126/science.abm8386} {\bibfield  {journal} {\bibinfo
  {journal} {Science}\ }\textbf {\bibinfo {volume} {375}},\ \bibinfo {pages}
  {774} (\bibinfo {year} {2022})}\BibitemShut {NoStop}%
\bibitem [{\citenamefont {Zhang}\ \emph {et~al.}(2023)\citenamefont {Zhang},
  \citenamefont {Polski}, \citenamefont {Thomson}, \citenamefont
  {Lantagne-Hurtubise}, \citenamefont {Lewandowski}, \citenamefont {Zhou},
  \citenamefont {Watanabe}, \citenamefont {Taniguchi}, \citenamefont {Alicea},\
  and\ \citenamefont {Nadj-Perge}}]{ZhangSC2023}%
  \BibitemOpen
  \bibfield  {author} {\bibinfo {author} {\bibfnamefont {Y.}~\bibnamefont
  {Zhang}}, \bibinfo {author} {\bibfnamefont {R.}~\bibnamefont {Polski}},
  \bibinfo {author} {\bibfnamefont {A.}~\bibnamefont {Thomson}}, \bibinfo
  {author} {\bibfnamefont {E}~\bibnamefont {Lantagne-Hurtubise}}, \bibinfo
  {author} {\bibfnamefont {C.}~\bibnamefont {Lewandowski}}, \bibinfo {author}
  {\bibfnamefont {H.}~\bibnamefont {Zhou}}, \bibinfo {author} {\bibfnamefont
  {K.}~\bibnamefont {Watanabe}}, \bibinfo {author} {\bibfnamefont
  {T.}~\bibnamefont {Taniguchi}}, \bibinfo {author} {\bibfnamefont
  {J.}~\bibnamefont {Alicea}}, \ and\ \bibinfo {author} {\bibfnamefont
  {S.}~\bibnamefont {Nadj-Perge}},\ }\href {\doibase
10.1038/s41586-022-05446-x} {\bibfield  {journal} {\bibinfo  {journal}
  {Nature}\ }\textbf {\bibinfo {volume} {613}},\ \bibinfo {pages} {268}
  (\bibinfo {year} {2023})}\BibitemShut {NoStop}%
\bibitem [{\citenamefont {Perali}\ \emph {et~al.}(2013)\citenamefont {Perali},
  \citenamefont {Neilson},\ and\ \citenamefont
  {Hamilton}}]{PhysRevLett.110.146803}%
  \BibitemOpen
  \bibfield  {author} {\bibinfo {author} {\bibfnamefont {A.}~\bibnamefont
  {Perali}}, \bibinfo {author} {\bibfnamefont {D.}~\bibnamefont {Neilson}}, \
  and\ \bibinfo {author} {\bibfnamefont {A.~R.}\ \bibnamefont {Hamilton}},\
  }\href {\doibase 10.1103/PhysRevLett.110.146803} {\bibfield  {journal}
  {\bibinfo  {journal} {Phys. Rev. Lett.}\ }\textbf {\bibinfo {volume} {110}},\
  \bibinfo {pages} {146803} (\bibinfo {year} {2013})}\BibitemShut {NoStop}%
\bibitem [{\citenamefont {Kosterlitz}\ and\ \citenamefont
  {Thouless}(1973)}]{KTtransition}%
  \BibitemOpen
  \bibfield  {author} {\bibinfo {author} {\bibfnamefont {J.~M.}\ \bibnamefont
  {Kosterlitz}}\ and\ \bibinfo {author} {\bibfnamefont {D.~J.}\ \bibnamefont
  {Thouless}},\ }\href {\doibase 10.1088/0022-3719/6/7/010} {\bibfield
  {journal} {\bibinfo  {journal} {Journal of Physics C: Solid State Physics}\
  }\textbf {\bibinfo {volume} {6}},\ \bibinfo {pages} {1181} (\bibinfo {year}
  {1973})}\BibitemShut {NoStop}%
\bibitem [{\citenamefont {Provost}\ and\ \citenamefont
  {Vallee}(1980)}]{provost1980riemannian}%
  \BibitemOpen
  \bibfield  {author} {\bibinfo {author} {\bibfnamefont {J.}~\bibnamefont
  {Provost}}\ and\ \bibinfo {author} {\bibfnamefont {G.}~\bibnamefont
  {Vallee}},\ }\href@noop {} {\bibfield  {journal} {\bibinfo  {journal}
  {Communications in Mathematical Physics}\ }\textbf {\bibinfo {volume} {76}},\
  \bibinfo {pages} {289} (\bibinfo {year} {1980})}\BibitemShut {NoStop}%
\bibitem [{\citenamefont {Chou}\ \emph {et~al.}(2022)\citenamefont {Chou},
  \citenamefont {Wu}, \citenamefont {Sau},\ and\ \citenamefont
  {Das~Sarma}}]{PhysRevB.105.L100503}%
  \BibitemOpen
  \bibfield  {author} {\bibinfo {author} {\bibfnamefont {Y.-Z.}\ \bibnamefont
  {Chou}}, \bibinfo {author} {\bibfnamefont {F.}~\bibnamefont {Wu}}, \bibinfo
  {author} {\bibfnamefont {J.~D.}\ \bibnamefont {Sau}}, \ and\ \bibinfo
  {author} {\bibfnamefont {S.}~\bibnamefont {Das~Sarma}},\ }\href {\doibase
  10.1103/PhysRevB.105.L100503} {\bibfield  {journal} {\bibinfo  {journal}
  {Phys. Rev. B}\ }\textbf {\bibinfo {volume} {105}},\ \bibinfo {pages}
  {L100503} (\bibinfo {year} {2022})}\BibitemShut {NoStop}%
\bibitem [{\citenamefont {Dong}\ \emph {et~al.}(2022)\citenamefont {Dong},
  \citenamefont {Chubukov},\ and\ \citenamefont {Levitov}}]{ChubukovLevitovSC}%
  \BibitemOpen
  \bibfield  {author} {\bibinfo {author} {\bibfnamefont {Z.}~\bibnamefont
  {Dong}}, \bibinfo {author} {\bibfnamefont {A.~V.}\ \bibnamefont {Chubukov}},
  \ and\ \bibinfo {author} {\bibfnamefont {L.}~\bibnamefont {Levitov}},\
  }\href@noop {} {\enquote {\bibinfo {title} {Spin-triplet superconductivity at
  the onset of isospin order in biased bilayer graphene},}\ } (\bibinfo {year}
  {2022}),\ \Eprint {http://arxiv.org/abs/2205.13353} {arXiv:2205.13353
  [cond-mat.supr-con]} \BibitemShut {NoStop}%
\bibitem [{\citenamefont {Eisenstein}\ and\ \citenamefont
  {MacDonald}(2004)}]{Eisenstein2004}%
  \BibitemOpen
  \bibfield  {author} {\bibinfo {author} {\bibfnamefont {J.~P.}\ \bibnamefont
  {Eisenstein}}\ and\ \bibinfo {author} {\bibfnamefont {A.~H.}\ \bibnamefont
  {MacDonald}},\ }\href {\doibase 10.1038/nature03081} {\bibfield  {journal}
  {\bibinfo  {journal} {Nature}\ }\textbf {\bibinfo {volume} {432}},\ \bibinfo
  {pages} {691} (\bibinfo {year} {2004})}\BibitemShut {NoStop}%
\bibitem [{\citenamefont {Blatt}\ \emph {et~al.}(1962)\citenamefont {Blatt},
  \citenamefont {B\"oer},\ and\ \citenamefont {Brandt}}]{PhysRev.126.1691}%
  \BibitemOpen
  \bibfield  {author} {\bibinfo {author} {\bibfnamefont {J.~M.}\ \bibnamefont
  {Blatt}}, \bibinfo {author} {\bibfnamefont {K.~W.}\ \bibnamefont {B\"oer}}, \
  and\ \bibinfo {author} {\bibfnamefont {W.}~\bibnamefont {Brandt}},\ }\href
  {\doibase 10.1103/PhysRev.126.1691} {\bibfield  {journal} {\bibinfo
  {journal} {Phys. Rev.}\ }\textbf {\bibinfo {volume} {126}},\ \bibinfo {pages}
  {1691} (\bibinfo {year} {1962})}\BibitemShut {NoStop}%
\bibitem [{\citenamefont {Min}\ \emph {et~al.}(2008)\citenamefont {Min},
  \citenamefont {Bistritzer}, \citenamefont {Su},\ and\ \citenamefont
  {MacDonald}}]{PhysRevB.78.121401}%
  \BibitemOpen
  \bibfield  {author} {\bibinfo {author} {\bibfnamefont {H.}~\bibnamefont
  {Min}}, \bibinfo {author} {\bibfnamefont {R.}~\bibnamefont {Bistritzer}},
  \bibinfo {author} {\bibfnamefont {J.-J.}\ \bibnamefont {Su}}, \ and\ \bibinfo
  {author} {\bibfnamefont {A.~H.}\ \bibnamefont {MacDonald}},\ }\href {\doibase
  10.1103/PhysRevB.78.121401} {\bibfield  {journal} {\bibinfo  {journal} {Phys.
  Rev. B}\ }\textbf {\bibinfo {volume} {78}},\ \bibinfo {pages} {121401}
  (\bibinfo {year} {2008})}\BibitemShut {NoStop}%
\bibitem [{\citenamefont {Zhang}\ and\ \citenamefont
  {Joglekar}(2008)}]{PhysRevB.77.233405}%
  \BibitemOpen
  \bibfield  {author} {\bibinfo {author} {\bibfnamefont {C.-H.}\ \bibnamefont
  {Zhang}}\ and\ \bibinfo {author} {\bibfnamefont {Y.~N.}\ \bibnamefont
  {Joglekar}},\ }\href {\doibase 10.1103/PhysRevB.77.233405} {\bibfield
  {journal} {\bibinfo  {journal} {Phys. Rev. B}\ }\textbf {\bibinfo {volume}
  {77}},\ \bibinfo {pages} {233405} (\bibinfo {year} {2008})}\BibitemShut
  {NoStop}%
\bibitem [{\citenamefont {Lutchyn}\ \emph {et~al.}(2010)\citenamefont
  {Lutchyn}, \citenamefont {Rossi},\ and\ \citenamefont
  {Das~Sarma}}]{lutchyn2010d}%
  \BibitemOpen
  \bibfield  {author} {\bibinfo {author} {\bibfnamefont {R.~M.}\ \bibnamefont
  {Lutchyn}}, \bibinfo {author} {\bibfnamefont {E.}~\bibnamefont {Rossi}}, \
  and\ \bibinfo {author} {\bibfnamefont {S.}~\bibnamefont {Das~Sarma}},\ }\href
  {\doibase 10.1103/PhysRevA.82.061604} {\bibfield  {journal} {\bibinfo
  {journal} {Physical Review A}\ }\textbf {\bibinfo {volume} {82}},\ \bibinfo
  {pages} {061604} (\bibinfo {year} {2010})}\BibitemShut {NoStop}%
\bibitem [{\citenamefont {Bistritzer}\ and\ \citenamefont
  {MacDonald}(2008)}]{PhysRevLett.101.256406}%
  \BibitemOpen
  \bibfield  {author} {\bibinfo {author} {\bibfnamefont {R.}~\bibnamefont
  {Bistritzer}}\ and\ \bibinfo {author} {\bibfnamefont {A.~H.}\ \bibnamefont
  {MacDonald}},\ }\href {\doibase 10.1103/PhysRevLett.101.256406} {\bibfield
  {journal} {\bibinfo  {journal} {Phys. Rev. Lett.}\ }\textbf {\bibinfo
  {volume} {101}},\ \bibinfo {pages} {256406} (\bibinfo {year}
  {2008})}\BibitemShut {NoStop}%
\bibitem [{\citenamefont {Zhang}\ and\ \citenamefont
  {Rossi}(2013)}]{PhysRevLett.111.086804}%
  \BibitemOpen
  \bibfield  {author} {\bibinfo {author} {\bibfnamefont {J.}~\bibnamefont
  {Zhang}}\ and\ \bibinfo {author} {\bibfnamefont {E.}~\bibnamefont {Rossi}},\
  }\href {\doibase 10.1103/PhysRevLett.111.086804} {\bibfield  {journal}
  {\bibinfo  {journal} {Phys. Rev. Lett.}\ }\textbf {\bibinfo {volume} {111}},\
  \bibinfo {pages} {086804} (\bibinfo {year} {2013})}\BibitemShut {NoStop}%
\bibitem [{\citenamefont {Kharitonov}\ and\ \citenamefont
  {Efetov}(2008)}]{PhysRevB.78.241401}%
  \BibitemOpen
  \bibfield  {author} {\bibinfo {author} {\bibfnamefont {M.~Y.}\ \bibnamefont
  {Kharitonov}}\ and\ \bibinfo {author} {\bibfnamefont {K.~B.}\ \bibnamefont
  {Efetov}},\ }\href {\doibase 10.1103/PhysRevB.78.241401} {\bibfield
  {journal} {\bibinfo  {journal} {Phys. Rev. B}\ }\textbf {\bibinfo {volume}
  {78}},\ \bibinfo {pages} {241401} (\bibinfo {year} {2008})}\BibitemShut
  {NoStop}%
\bibitem [{\citenamefont {Kharitonov}\ and\ \citenamefont
  {Efetov}(2010)}]{Kharitonov_2010}%
  \BibitemOpen
  \bibfield  {author} {\bibinfo {author} {\bibfnamefont {M.~Y.}\ \bibnamefont
  {Kharitonov}}\ and\ \bibinfo {author} {\bibfnamefont {K.~B.}\ \bibnamefont
  {Efetov}},\ }\href {\doibase 10.1088/0268-1242/25/3/034004} {\bibfield
  {journal} {\bibinfo  {journal} {Semiconductor Science and Technology}\
  }\textbf {\bibinfo {volume} {25}},\ \bibinfo {pages} {034004} (\bibinfo
  {year} {2010})}\BibitemShut {NoStop}%
\bibitem [{\citenamefont {Bistritzer}\ \emph {et~al.}(2008)\citenamefont
  {Bistritzer}, \citenamefont {Min}, \citenamefont {Su},\ and\ \citenamefont
  {MacDonald}}]{bistritzer2008commentelectronscreeningexcitonic}%
  \BibitemOpen
  \bibfield  {author} {\bibinfo {author} {\bibfnamefont {R.}~\bibnamefont
  {Bistritzer}}, \bibinfo {author} {\bibfnamefont {H.}~\bibnamefont {Min}},
  \bibinfo {author} {\bibfnamefont {J.~J.}\ \bibnamefont {Su}}, \ and\ \bibinfo
  {author} {\bibfnamefont {A.~H.}\ \bibnamefont {MacDonald}},\ }\href
  {https://arxiv.org/abs/0810.0331} {\  (\bibinfo {year} {2008})},\ \Eprint
  {http://arxiv.org/abs/0810.0331} {arXiv:0810.0331 [cond-mat.mes-hall]}
  \BibitemShut {NoStop}%
\bibitem [{\citenamefont {Sodemann}\ \emph {et~al.}(2012)\citenamefont
  {Sodemann}, \citenamefont {Pesin},\ and\ \citenamefont
  {MacDonald}}]{PhysRevB.85.195136}%
  \BibitemOpen
  \bibfield  {author} {\bibinfo {author} {\bibfnamefont {I.}~\bibnamefont
  {Sodemann}}, \bibinfo {author} {\bibfnamefont {D.~A.}\ \bibnamefont {Pesin}},
  \ and\ \bibinfo {author} {\bibfnamefont {A.~H.}\ \bibnamefont {MacDonald}},\
  }\href {\doibase 10.1103/PhysRevB.85.195136} {\bibfield  {journal} {\bibinfo
  {journal} {Phys. Rev. B}\ }\textbf {\bibinfo {volume} {85}},\ \bibinfo
  {pages} {195136} (\bibinfo {year} {2012})}\BibitemShut {NoStop}%
\bibitem [{\citenamefont {Abergel}\ \emph {et~al.}(2013)\citenamefont
  {Abergel}, \citenamefont {Rodriguez-Vega}, \citenamefont {Rossi},\ and\
  \citenamefont {Das~Sarma}}]{PhysRevB.88.235402}%
  \BibitemOpen
  \bibfield  {author} {\bibinfo {author} {\bibfnamefont {D.~S.~L.}\
  \bibnamefont {Abergel}}, \bibinfo {author} {\bibfnamefont {M.}~\bibnamefont
  {Rodriguez-Vega}}, \bibinfo {author} {\bibfnamefont {E.}~\bibnamefont
  {Rossi}}, \ and\ \bibinfo {author} {\bibfnamefont {S.}~\bibnamefont
  {Das~Sarma}},\ }\href {\doibase 10.1103/PhysRevB.88.235402} {\bibfield
  {journal} {\bibinfo  {journal} {Phys. Rev. B}\ }\textbf {\bibinfo {volume}
  {88}},\ \bibinfo {pages} {235402} (\bibinfo {year} {2013})}\BibitemShut
  {NoStop}%
\bibitem [{\citenamefont {Scammell}\ and\ \citenamefont
  {Sushkov}(2023)}]{PhysRevResearch.5.043176}%
  \BibitemOpen
  \bibfield  {author} {\bibinfo {author} {\bibfnamefont {H.~D.}\ \bibnamefont
  {Scammell}}\ and\ \bibinfo {author} {\bibfnamefont {O.~P.}\ \bibnamefont
  {Sushkov}},\ }\href {\doibase 10.1103/PhysRevResearch.5.043176} {\bibfield
  {journal} {\bibinfo  {journal} {Phys. Rev. Res.}\ }\textbf {\bibinfo {volume}
  {5}},\ \bibinfo {pages} {043176} (\bibinfo {year} {2023})}\BibitemShut
  {NoStop}%
\bibitem [{\citenamefont {Su}\ and\ \citenamefont
  {MacDonald}(2017)}]{PhysRevB.95.045416}%
  \BibitemOpen
  \bibfield  {author} {\bibinfo {author} {\bibfnamefont {J.-J.}\ \bibnamefont
  {Su}}\ and\ \bibinfo {author} {\bibfnamefont {A.~H.}\ \bibnamefont
  {MacDonald}},\ }\href {\doibase 10.1103/PhysRevB.95.045416} {\bibfield
  {journal} {\bibinfo  {journal} {Phys. Rev. B}\ }\textbf {\bibinfo {volume}
  {95}},\ \bibinfo {pages} {045416} (\bibinfo {year} {2017})}\BibitemShut
  {NoStop}%
\bibitem [{\citenamefont {Sreejith}\ \emph {et~al.}(2024)\citenamefont
  {Sreejith}, \citenamefont {Sau},\ and\ \citenamefont
  {Das~Sarma}}]{PhysRevLett.133.056501}%
  \BibitemOpen
  \bibfield  {author} {\bibinfo {author} {\bibfnamefont {G.~J.}\ \bibnamefont
  {Sreejith}}, \bibinfo {author} {\bibfnamefont {J.~D.}\ \bibnamefont {Sau}}, \
  and\ \bibinfo {author} {\bibfnamefont {S.}~\bibnamefont {Das~Sarma}},\ }\href
  {\doibase 10.1103/PhysRevLett.133.056501} {\bibfield  {journal} {\bibinfo
  {journal} {Phys. Rev. Lett.}\ }\textbf {\bibinfo {volume} {133}},\ \bibinfo
  {pages} {056501} (\bibinfo {year} {2024})}\BibitemShut {NoStop}%
\bibitem [{\citenamefont {Neilson}\ \emph {et~al.}(2014)\citenamefont
  {Neilson}, \citenamefont {Perali},\ and\ \citenamefont
  {Hamilton}}]{PhysRevB.89.060502}%
  \BibitemOpen
  \bibfield  {author} {\bibinfo {author} {\bibfnamefont {D.}~\bibnamefont
  {Neilson}}, \bibinfo {author} {\bibfnamefont {A.}~\bibnamefont {Perali}}, \
  and\ \bibinfo {author} {\bibfnamefont {A.~R.}\ \bibnamefont {Hamilton}},\
  }\href {\doibase 10.1103/PhysRevB.89.060502} {\bibfield  {journal} {\bibinfo
  {journal} {Phys. Rev. B}\ }\textbf {\bibinfo {volume} {89}},\ \bibinfo
  {pages} {060502} (\bibinfo {year} {2014})}\BibitemShut {NoStop}%
\bibitem [{\citenamefont {Debnath}\ \emph {et~al.}(2017)\citenamefont
  {Debnath}, \citenamefont {Barlas}, \citenamefont {Wickramaratne},
  \citenamefont {Neupane},\ and\ \citenamefont {Lake}}]{PhysRevB.96.174504}%
  \BibitemOpen
  \bibfield  {author} {\bibinfo {author} {\bibfnamefont {B.}~\bibnamefont
  {Debnath}}, \bibinfo {author} {\bibfnamefont {Y.}~\bibnamefont {Barlas}},
  \bibinfo {author} {\bibfnamefont {D.}~\bibnamefont {Wickramaratne}}, \bibinfo
  {author} {\bibfnamefont {M.~R.}\ \bibnamefont {Neupane}}, \ and\ \bibinfo
  {author} {\bibfnamefont {R.~K.}\ \bibnamefont {Lake}},\ }\href {\doibase
  10.1103/PhysRevB.96.174504} {\bibfield  {journal} {\bibinfo  {journal} {Phys.
  Rev. B}\ }\textbf {\bibinfo {volume} {96}},\ \bibinfo {pages} {174504}
  (\bibinfo {year} {2017})}\BibitemShut {NoStop}%
\end{thebibliography}

%

\end{document}